\documentclass[sigplan,screen]{acmart}

\AtBeginDocument{%
  }

\copyrightyear{2025}
\acmYear{2025}
\setcopyright{cc}
\setcctype{by}
\acmConference[SOSP '25]{ACM SIGOPS 31st Symposium on Operating Systems Principles}{October 13--16, 2025}{Seoul, Republic of Korea}
\acmBooktitle{ACM SIGOPS 31st Symposium on Operating Systems Principles (SOSP '25), October 13--16, 2025, Seoul, Republic of Korea}\acmDOI{10.1145/3731569.3764839}
\acmISBN{979-8-4007-1870-0/2025/10}

\usepackage[usenames,dvipsnames]{xcolor,colortbl}
\usepackage{amsmath}
\usepackage[utf8]{inputenc}
\usepackage{amsmath,amsthm}
\usepackage{MnSymbol}
\usepackage{cleveref}
\usepackage{listings, fancyvrb, multicol}
\usepackage{enumitem}
\usepackage{float}
\usepackage{tikz-cd}
\usepackage{microtype}
\usepackage{listings}
\usepackage{comment}
\usepackage{xspace}
\usepackage{graphicx, graphics}
\usepackage{subcaption}
\usepackage{svg}
\usepackage{balance}
\usepackage{etoolbox}
\newcommand{\sailor}{{Sailor}\xspace}

\newcommand{\new}[1]{{#1~}}

\newcommand*\circled[1]{\tikz[baseline=(char.base)]{
            \node[shape=circle,draw,inner sep=2pt] (char) {#1};}}

\begin{document}

\title[\sailor: Automating Distributed Training over Dynamic, \\ Heterogeneous, and Geo-distributed Clusters]{\sailor: Automating Distributed Training over Dynamic, Heterogeneous, and Geo-distributed Clusters}
\settopmatter{authorsperrow=5}

\author{Foteini Strati}
\authornote{Correspondence to \href{foteini.strati@inf.ethz.ch}{foteini.strati@inf.ethz.ch}}
\affiliation{%
  \institution{ETH Zurich}
  \country{Switzerland}
}

\author{Zhendong Zhang}
\affiliation{%
  \institution{ETH Zurich}
  \country{Switzerland}
}

\author{George Manos}
\affiliation{%
  \institution{ETH Zurich}
  \country{Switzerland}
}

\author{Ixeia Sánchez Périz}
\authornote{Work done while at ETH Zurich}
\affiliation{%
  \institution{}
  \country{}
}

\author{Qinghao Hu}
\affiliation{%
  \institution{MIT}
   \country{USA}
}

\author{Tiancheng Chen}
\affiliation{%
  \institution{ETH Zurich}
  \country{Switzerland}
}

\author{Berk Buzcu}
\affiliation{%
  \institution{HES-SO}
  \country{Switzerland}
}

\author{Song Han}
\affiliation{%
  \institution{MIT}
  \country{USA}
}

\author{Pamela Delgado}
\affiliation{%
  \institution{HES-SO}
  \country{Switzerland}
}

\author{Ana Klimovic}
\affiliation{%
  \institution{ETH Zurich}
  \country{Switzerland}
}

\renewcommand{\shortauthors}{Strati, et al.}

\begin{abstract}
The high GPU demand of ML training makes it hard to allocate large homogeneous clusters of high-end GPUs in a single availability zone. Leveraging  heterogeneous GPUs available within and across zones can improve throughput at a reasonable cost. However, training ML models on heterogeneous resources introduces significant challenges, such as stragglers and a large search space of possible job configurations. Current systems lack support for efficiently training models on heterogeneous resources. We present \sailor, a system that automates distributed training over heterogeneous, geo-distributed, and dynamically available resources. \sailor combines an efficient search space exploration algorithm, accurate runtime and memory footprint simulation, and a distributed training framework that supports different types of heterogeneity to optimize training throughput and cost.
\end{abstract}


\maketitle

\section{Introduction}\label{intro}

GPUs are in high demand for large-scale Machine Learning (ML). As ML models continue to grow exponentially in size, they require an increasing number of GPUs to train and fine-tune. This high demand makes it difficult for model developers to allocate the desired number of accelerators to train models at high throughput in public clouds or enterprise clusters~\cite{Um24Metis, strati24crossregion, Zongheng23Skypilot}. 
Datacenters typically host a variety of GPU types and generations, spread across geographic regions~\cite{gcpgpus, Jia2022Whale, weng22mlaas, Jeon19philly, wu2024falcon}. Yet model developers tend to restrict model training to \textit{homogeneous} clusters of GPUs, since state-of-the-art distributed training frameworks like Megatron-LM~\cite{shoeybi2020megatronlm} and DeepSpeed~\cite{megatron-deepspeed} assume homogeneous GPUs and inter-node bandwidth. The demand for large, homogeneous GPU clusters compounds the scarcity of high-end GPUs.

\begin{figure}[t]
\centering
  \includegraphics[width=\linewidth]{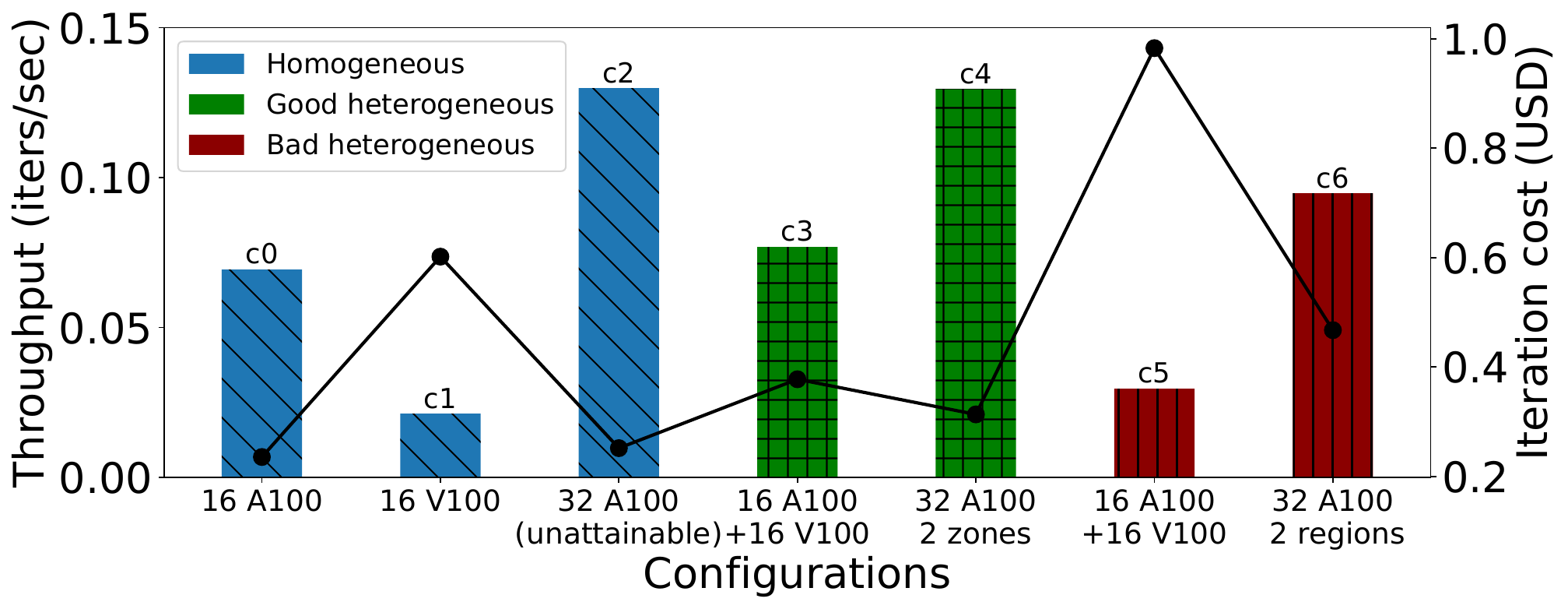}\label{fig:motivation_het}%
\vspace{-5pt}
\caption{When homogeneous resources are limited, using heterogeneous GPUs (A100, V100) or multiple zones can increase OPT-350M training throughput (c3-c4) at low cost (black line). However, when the resource topology and job parallelization are not well selected, iteration time and monetary cost may increase significantly (c5-c6).} 
\vspace{-3pt}
\label{fig:motivation_het}
\end{figure}

Allowing a training job to run on heterogeneous GPU types and/or GPUs distributed across zones (i.e., with heterogeneous inter-node bandwidth) can give model developers access to \textit{more GPUs} per job to increase training throughput. 
For example, consider a model developer who seeks to maximize training throughput by using 32 A100 GPUs (\textcolor{blue}{\texttt{c2}} in Figure~\ref{fig:motivation_het}), but discovers that only 16 A100s (\textcolor{blue}{\texttt{c0}} in Figure~\ref{fig:motivation_het}) are currently available in one zone. Using heterogeneous GPU generations (e.g., \textcolor{teal}{\texttt{c3}} uses an additional 16 V100s in the zone) or multi-zone configurations (e.g., \textcolor{teal}{\texttt{c4}} uses 32 A100s spread across 2 zones within a region) increases throughput by 1.15$\times$ and 1.87$\times$, respectively, with a moderate increase in monetary cost per iteration (black line).

However, supporting heterogeneous, geo-distributed resources introduces several challenges.
First, heterogeneous GPU types and placements across zones exponentially expand the configuration search space. 
Searching for an optimal configuration requires jointly optimizing the \textit{resource allocation} and \textit{job parallelization plan} (e.g., data/pipeline/tensor parallelism degrees). For example, in Figure~\ref{fig:motivation_het}, although \textcolor{red}{\texttt{c5}} uses the same number of GPUs as \textcolor{teal}{\texttt{c3}}, it achieves much lower throughput due to a suboptimal parallelization plan. 
The search must also consider the cost caused by extra resources and data transfers. Cloud providers charge significant fees for inter-zone and inter-region data transfers~\cite{aws-interzone, gcp-interzone, az-interzone}, which impacts geo-distributed configurations. In Figure~\ref{fig:motivation_het}, \textcolor{red}{\texttt{c6}} uses the same number of GPUs and parallelization strategy as \textcolor{teal}{\texttt{c4}}, but spreads training across regions (instead of zones within the same region), which increases cost. 

Furthermore, as resource availability changes dynamically in datacenters, due to variable demand and node failures or preemptions, it is necessary to navigate this vast search space quickly ~\cite{Thorpe23Bamboo, wagenlander24tenplex}. Figure~\ref{fig:trace} shows the varying number of A100 GPUs that we were able to allocate over an 8-hour period, out of the 8 A100s that we continuously requested in 2 different zones in Google Cloud. Such changes in resource availability require frequently re-evaluating the job configuration search space. Thus, model developers require a system to quickly navigate the large search space of heterogeneous (and homogeneous) configurations with dynamic resource availability, optimizing for the user's performance-cost objective, while satisfying constraints (e.g., budget limits).

Second, profiling many candidate configurations to evaluate their throughput is prohibitively expensive and time-consuming. Hence, it is critical to accurately estimate the iteration time of a candidate configuration. This is challenging in heterogeneous environments, as differences in peak FLOPS, memory capacity, CPU-GPU interconnects, number of GPUs per node, and inter-node bandwidth typically lead to \textit{stragglers}, which can significantly limit throughput. 
More importantly, variability in memory capacity per GPU may cause \textit{out-of-memory (OOM) errors} in some GPUs, disrupting the entire training job. Simulating iteration time and checking that job configurations are valid (i.e., will not cause OOM) requires correctly modeling stragglers and per-GPU memory footprints.

Finally, after finding an appropriate resource allocation and parallelization plan for a training job given the available resources, model developers need to be able to run this job configuration in a distributed training framework (such as Megatron~\cite{shoeybi2020megatronlm}). We find that optimal configurations for jobs running on heterogeneous resource topologies often include heterogeneous parallelism degrees per stage to load-balance the compute and memory capacity of different GPU types. 
Today's state-of-the-art distributed training frameworks need to be adapted to support such heterogeneous job configurations. Furthermore, as resource availability can change frequently (e.g., when using spot instances~\cite{Thorpe23Bamboo}), the training framework must be able to quickly reconfigure jobs.

Existing systems do not adequately solve these challenges. First,   
current works do not co-optimize the resource allocation with the job parallelization plan. Instead, systems like Aceso~\cite{Liu24Aceso}, Galvatron~\cite{Miao22Galvatron}, and others in Table~\ref{table:planners} expect the user to select a fixed resource allocation for which the system recommends a job parallelization plan. Most systems also do not consider heterogeneous resource topologies. 
Recent systems like Atlas~\cite{palak2024atlas}, DTFM~\cite{yuan2023decentralized}, Metis~\cite{Um24Metis}, and FlashFlex~\cite{yan2024flashflex} optimize parallelization for heterogeneous GPUs or geo-distributed setups, but they suffer from prohibitively long search times (up to hours for configurations with 10s of GPUs)~\cite{Um24Metis}, or suboptimal cost functions~\cite{yan2024flashflex, yuan2023decentralized}, making them unsuitable for environments with dynamic resource availability. 
Second, existing systems rely on inaccurate simulators to estimate the training throughput and memory footprint of candidate configurations. For example, Varuna~\cite{Athlur22Varuna} overlooks significant memory sources (e.g., memory needed by the optimizer, communication, etc) when estimating memory footprint, hence recommending configurations that lead to OOM errors. Finally, state-of-the-art distributed training frameworks like Megatron-LM~\cite{shoeybi2020megatronlm} are slow to reconfigure jobs and do not support heterogeneous job parallelization plans or different microbatch sizes per GPU, which is necessary to maximize throughput and minimize cost in heterogeneous clusters.

To this end, we propose \textit{\sailor}, a system for efficient large-scale training over heterogeneous resources with dynamic availability. \sailor\footnote{\sailor is available at \href{https://github.com/eth-easl/sailor}{https://github.com/eth-easl/sailor}} consists of three components: a configuration planner, a simulator, and a distributed training framework. The \sailor planner navigates the search space of resource allocations and job parallelization plan combinations. It recommends configurations that optimize a user-defined objective (e.g., max throughput or min cost) under constraints (e.g., max budget or min throughput). The planner considers heterogeneous GPU and machine types and geo-distributed setups. The planner uses the simulator to accurately model iteration time and memory footprint for any given configuration. 
Through a combination of dynamic programming and search space pruning with effective heuristics, the planner finds solutions within seconds even for allocations with 100s of GPUs and varying degrees of heterogeneity. This allows \sailor to quickly adapt plans based on resource availability. Finally, the \sailor training framework adds support for heterogeneous configurations to execute the planner’s configurations. It also adds support for fault tolerance and elasticity, enabling adaptation to changes in resource availability. Together, these components enable \sailor to efficiently automate large-scale training in homogeneous, heterogeneous, and/or dynamic resource environments.

We evaluate \sailor in various setups and compare it extensively to prior works. 
To the best of our knowledge, our work is the first to compare the major open-source ML training planners proposed to-date (Table \ref{table:planners}) in homogeneous and heterogeneous scenarios. We show that \sailor can find resource allocations and job parallelization plans that result in higher throughput than baselines in both homogeneous and heterogeneous clusters.
We show how \sailor can leverage heterogeneous resources to improve throughput by 1.1-2.87$\times$ compared to the heterogeneity-aware baselines (Metis, FlashFlex, AMP), while maintaining search times of 10s of seconds compared to minutes or hours needed by the baselines. We also show \sailor's ability to increase performance and reduce cost in geo-distributed setups compared to DTFM by 5.9$\times$ and 9.8$\times$, respectively. Finally, we demonstrate \sailor's ability to minimize monetary cost given throughput constraints, resulting in 40\% cost savings compared to the second-best-performing baseline.
\section{Background}\label{sec:background}

\subsection{ML job parallelization strategies}

\begin{table}[tp!] 
\footnotesize
\centering
\begin{tabular}{|l|c|c|}
\hline
\textbf{System} & \textbf{Support} & \textbf{Search Time (128 A100)} \\
\hline
Piper~\cite{Tarnawski21Piper} & 3D, X, X, X & <1 sec \\
\hline
AMP~\cite{li2022amp} & 3D, X, $\checkmark$, X & 14 sec  \\
\hline
Varuna~\cite{Athlur22Varuna} & 3D, X, X, X & < 1 sec  \\
\hline
Oobleck~\cite{Jang2023Oobleck} & 3D, X, X, X & hours \\
\hline
Metis~\cite{Um24Metis} & 3D, X, $\checkmark$, X & hours \\
\hline
FlashFlex~\cite{yan2024flashflex} & 3D, $\checkmark$, $\checkmark$, X & 3 sec  \\
\hline
Galvatron~\cite{Miao22Galvatron} & 3D, X, X, X  & 10s of sec \\
\hline
Aceso~\cite{Liu24Aceso} & 3D, X, X, X & 200 sec  \\
\hline
DTFM~\cite{yuan2023decentralized} & 2D, $\checkmark$, X, $\checkmark$ & 125 sec \\
\hline
Atlas~\cite{palak2024atlas} & 3D, $\checkmark$, X, $\checkmark$ & 100 sec \\
\hline
\textbf{\sailor} & \textbf{3D, $\checkmark$,$\checkmark$,$\checkmark$} & \textbf{< 1 sec}
\\
\hline
\end{tabular}

\caption{Overview of distributed ML training planners. We omit planners that change ML training semantics~\cite{Wang23cocktailsgd}. The \textit{Support} column stands for: [\textit{degrees of parallelism supported, recommends resource allocation, supports heterogeneous GPU types, supports multi-zone}].
Search time assumes a cluster of 128 A100 GPUs and OPT-350M model. 
}
\label{table:planners}
\vspace{-0.5cm}

\end{table}

Million or billion-parameter ML models train on massive datasets on clusters of high-end accelerators such as GPUs, using a combination of parallelization strategies:

\textbf{Data Parallelism (DP):} The model is replicated across workers, while the dataset is partitioned. At the end of each iteration, the workers synchronize their gradients using an all-reduce collective~\cite{Sapio21innetwork}.

\textbf{Pipeline Parallelism (PP):} The model is split into stages, with each stage consisting of a set of layers, and assigned to a worker or node. Workers operate at the granularity of a microbatch\footnote{A minibatch is split in microbatches.}, performing forward and backward passes, sending activations to the next stage, and gradients to the previous stage. Due to inter-stage dependencies, pipeline parallelism is subject to \textit{bubbles}, i.e., periods that a stage remains idle, waiting for others to complete~\cite{Huang19Gpipe}. As a result, many approaches have been proposed to process multiple microbatches simultaneously and reduce bubbles~\cite{Athlur22Varuna, Narayanan19Pipedream}.

\textbf{Tensor Parallelism (TP):} With tensor parallelism, a layer is divided across GPUs. After each GPU performs its local computations (both in forward and backward pass), the GPUs are synchronized using collectives such as all-reduce and all-gather. Since TP requires frequent communication, it requires very high interconnection bandwidths and is usually limited within a single node for reasonable throughput~\cite{shoeybi2020megatronlm}.

\vspace{-0.1cm}
\subsection{Automating parallelization strategies}

Determining the optimal degree of parallelism for each dimension (DP, PP, TP) is complex and greatly affects training throughput. Multiple systems, which we refer to as \textit{planners}, automate this process (see Table~\ref{table:planners}). Given a fixed resource allocation (e.g., 16 nodes with 4 A100 GPUs each), model configuration (including hyperparameters like global batch size and learning rate), model profiling information (e.g., time for forward and backward pass for different configurations), and hardware characteristics (e.g., network bandwidth), planners explore different parallelization strategies, estimating the training time under different configurations, using some form of simulation. Planners apply techniques such as exhaustive search~\cite{Athlur22Varuna, Um24Metis}, dynamic programming~\cite{Tarnawski21Piper, li2022amp}, and integer linear programming~\cite{zheng2022alpa} to identify configurations that minimize training time.

\section{Motivation and Challenges}\label{sec:motivation_challenges}

\begin{figure}[tp!]
\centering
  \includegraphics[width=\linewidth]{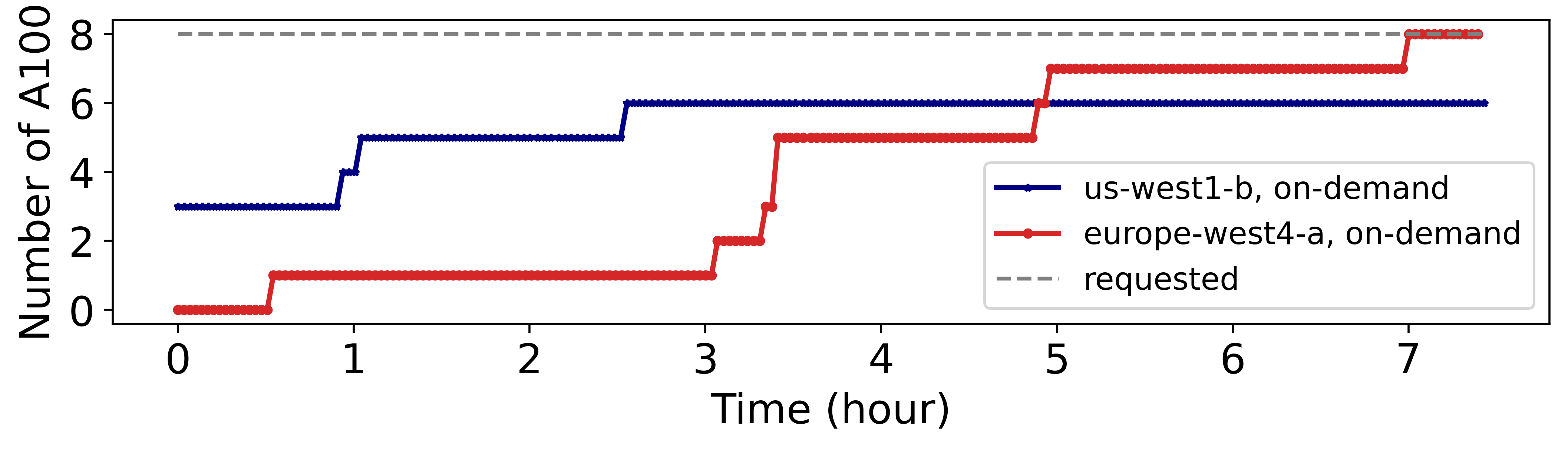}
\vspace{-15pt}
\caption{Availability of A100 GPUs in two zones in Google Cloud over 8-hr period. We request 8 GPUs in each zone. The trace was collected in April 2024.} 
\label{fig:trace}
\vspace{-0.5cm}
\end{figure}

\subsection{Why use heterogeneous, geo-distributed GPUs?}\label{motivation}

\textbf{Homogeneous high-end GPU clusters are scarce.} The widespread adoption of ML, and hardware vendors' inability to keep up with this pace, has significantly increased GPU demand. Several studies report limited GPU availability across public cloud providers~\cite{Zongheng23Skypilot, Cheng23availability, chugh23availability, strati24crossregion, Thorpe23Bamboo}. 
For example, Figure \ref{fig:trace} plots the availability of on-demand A100 GPUs in two zones in GCP. In one zone, it took 7 hours to allocate 8 A100 GPUs, while in the other zone, the requested number of GPUs was not attained within the 8-hour window.
Our findings align with the AWS GPU availability plot from Guo et al.~\cite{guo2024cephalo}, which shows that high-end GPUs such as A100 and H100 are difficult to aquire, while mid-tier GPUs (A10G, V100, T4) have higher but still limited and dynamic availability. 
Using GPUs across zones with heterogeneous types gives ML jobs the opportunity to use more GPUs to further increase throughput, as shown in Figure~\ref{fig:motivation_het}.

\textbf{Power and cooling limits \#GPUs per datacenter.}  From 2020 to 2025, the size of state-of-the-art ML models has increased by roughly 1200$\times$~\cite{gpt-2, gpt-4, llama-4}, while per-GPU memory capacity has only increased by 8$\times$~\cite{nvidia-list}. This requires hyperscalers to deploy more and more GPUs~\cite{ms-newAIdatacenter, meta-newAIdatacenter}. 
However, the high power and cooling requirements of high-end GPUs limits the amount that can be deployed per datacenter~\cite{tapas}. The next generation of ML models may need to train on GPUs across multiple availability zones~\cite{geodistributedblog1, geodistributedblog2, heatchallengeblog, palak2024improvingtrainingtimegpu}.

\textbf{Using old GPUs can reduce embodied carbon.} Embodied carbon (greenhouse gas emissions associated with manufacturing to disposal) is a major source of datacenter  emissions~\cite{greenSKU, carbonExplorer, metaEmbodiedCarbon}.
Although users prefer to allocate the latest GPUs for their ML jobs, older GPUs are abundant as the typical lifetime for ML servers in hyperscaler datacenters is $\sim$6 years~\cite{schneider2025lifecycleGoogleAI, greenSKU}. Finding optimal ways to spread jobs across heterogeneous GPUs will enable leveraging older GPUs for longer to better amortize embodied carbon.

\subsection{Challenges with Heterogeneous ML Clusters}\label{challenges}

\textbf{C1: Quickly searching a vast configuration space.} 
Considering heterogeneous and geo-distributed GPUs creates a vast and complex search space.
The ML developer needs to decide how many GPUs to use and how to group them across VMs.
This complexity increases further, when accounting for job parallelization plans within each allocation. Furthermore, the optimal allocation and partitioning strategies depend on user objectives and constraints.
A planner needs to \textit{quickly} navigate the large configuration space to adapt to dynamic resource availability in cloud and on-premise environments~\cite{Thorpe23Bamboo, guo2024cephalo, weng22mlaas, Gu19Tiresias}, since maximizing throughput requires adapting parallelization strategies with changes to cluster topology~\cite{Athlur22Varuna, wagenlander24tenplex}.

Table~\ref{table:planners} shows that Metis~\cite{Um24Metis}, FlashFlex~\cite{yan2024flashflex}, Cephalo~\cite{guo2024cephalo}, Atlas~\cite{palak2024atlas}, and DTFM~\cite{yuan2023decentralized}  explore parts of this large search space. Atlas~\cite{palak2024atlas} and DTFM~\cite{yuan2023decentralized}  target geo-distributed training, but do not consider heterogeneous GPU types, and do not decide the various parallelism degrees: instead, they take as input the parallelism degrees, and assign these degrees in the available zones.
On the other hand, Metis~\cite{Um24Metis}, FlashFlex~\cite{yan2024flashflex} and Cephalo~\cite{guo2024cephalo},
consider heterogeneous GPU types, but overlook geo-distributed training, and are quite inefficient for dynamic environments. Metis needs a few hours to devise a plan for a 16-GPU cluster (A100 and V100)\footnote{with max\_permutation\_length and device group variance set to 10 and 0.5 respectively, according to the paper~\cite{Um24Metis}}, making frequent reevaluation infeasible as GPU availability changes. Cephalo~\cite{guo2024cephalo} has a search time of ~300 sec on a cluster of 64 GPUs\footnote{reported on the paper~\cite{guo2024cephalo}}, but it is limited only to Fully Sharded Data Parallelism. FlashFlex~\cite{yan2024flashflex} has a short runtime, but provides inaccurate runtime estimations, leading to suboptimal plans. 
Furthermore, these planners only optimize for \textit{throughput}, ignoring budget constraints, and cost (dollars per iteration), which affect the optimal configuration (\S\ref{sec:eval_constraints}).

\textbf{C2: Accurately simulating memory footprint and iteration time.} Most planners use analytical models or simulations to evaluate a configuration (parallelism degrees, microbatch sizes, etc) on a given cluster setup, since it is impractical and very expensive to deploy and profile every configuration. This evaluation usually includes two stages: 1) memory footprint estimation, to identify whether a configuration is valid (i.e., it does not lead to OOM errors) and 2) iteration time estimation, to determine performance. 

{\setlength{\textfloatsep}{-0cm}
\begin{figure}[tp!]
\centering
\vspace{5pt}
  \includegraphics[width=\columnwidth]{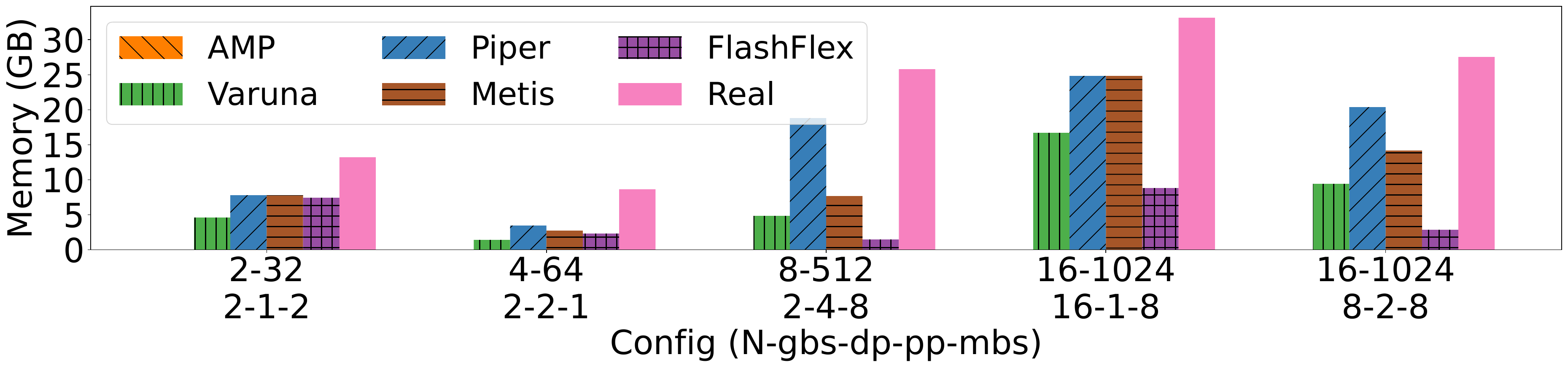}%
\caption{Peak Memory estimations of various baselines compared to the actual peak memory, for the OPT-350M model on a homogeneous cluster of 4 Grace-Hopper per node.  $N$ stands for the number of nodes,  $gbs$ is the global batch size, $pp$ pipeline parallelism, $tp$ tensor parallelism, $dp$ data parallelism, and $mbs$ the microbatch size. }
\label{fig:motivation_sim_mem}
\end{figure}}

Unfortunately, these estimations are often inaccurate, resulting in suboptimal or invalid plans. First,  planners either completely ignore memory footprint~\cite{li2022amp}, or underestimate the amount of memory requirements during training~\cite{Athlur22Varuna, Jang2023Oobleck}, omitting activations, optimizer states, and memory fragmentation, or assume the training memory footprint is uniform across all devices and pipeline stages~\cite{yan2024flashflex, Lin24nnScaler}. As a result, these systems may find plans that cause OOM errors, when deployed. Figure \ref{fig:motivation_sim_mem} compares memory footprint estimations with the real footprint on a homogeneous cluster of up to 16 Grace-Hopper nodes for the OPT-350M model, showing that planners can be 25-95\% off when estimating memory footprint.  Second, planners often poorly model training time, due to incorrect assumptions about network bandwidth and ignoring communication-computation overlap~\cite{Zhang17Poseidon, megatron-deepspeed}.

Modeling memory footprint and iteration time becomes even more complex with heterogeneity~\cite{Mo24Heet}. Different GPU generations vary in compute performance (e.g., TFLOPs), and memory capacity~\cite{zhu2024nanoflow}. Thus, a configuration that fits in one GPU, might cause OOM errors in another GPU. Additionally, stragglers and network bandwidth differences (especially in geo-distributed setups~\cite{strati24crossregion}) must be considered for accurate timing estimations. Planners must accurately model compute, memory, and network bandwidth heterogeneity - yet, as shown in Table~\ref{table:planners}, most systems overlook this.

\textbf{C3: Supporting both heterogeneous plans and seamless elasticity in a real distributed training framework.}   
Most training frameworks, i.e., systems that train a model on a set of devices, assume homogeneous clusters and job configuration plans. For example, widely used and high-performing frameworks such as Megatron~\cite{shoeybi2020megatronlm} and DeepSpeed~\cite{megatron-deepspeed} assume \textit{uniform} parallelism degrees for the entire training job (e.g. DP=2, PP=6, TP=1). This limits efficiency in heterogeneous configurations, where various parallelism degrees per training subproblem (e.g., using PP=6, with first 3 stages having TP=4, and the next 3 stages having TP=2) help load-balance compute and memory on GPU nodes with different resources. Thus, a framework should accommodate heterogeneous degrees of parallelism. Furthermore, the framework should seamlessly reconfigure and adapt to dynamic GPU availability (Figure \ref{fig:trace}), as long job reconfiguration times in response to resource changes are wasteful~\cite{Thorpe23Bamboo, wagenlander24tenplex}. Although related works propose elastic systems~\cite{Athlur22Varuna, Thorpe23Bamboo, Duan24Parcae} or systems that support heterogeneity~\cite{Miao23SDPipe, yan2024flashflex}, there is no open-source system that supports both.

\section{\sailor}\label{sec:sailor}

\begin{figure}[t!]
    \centering
    \includegraphics[width=\linewidth]{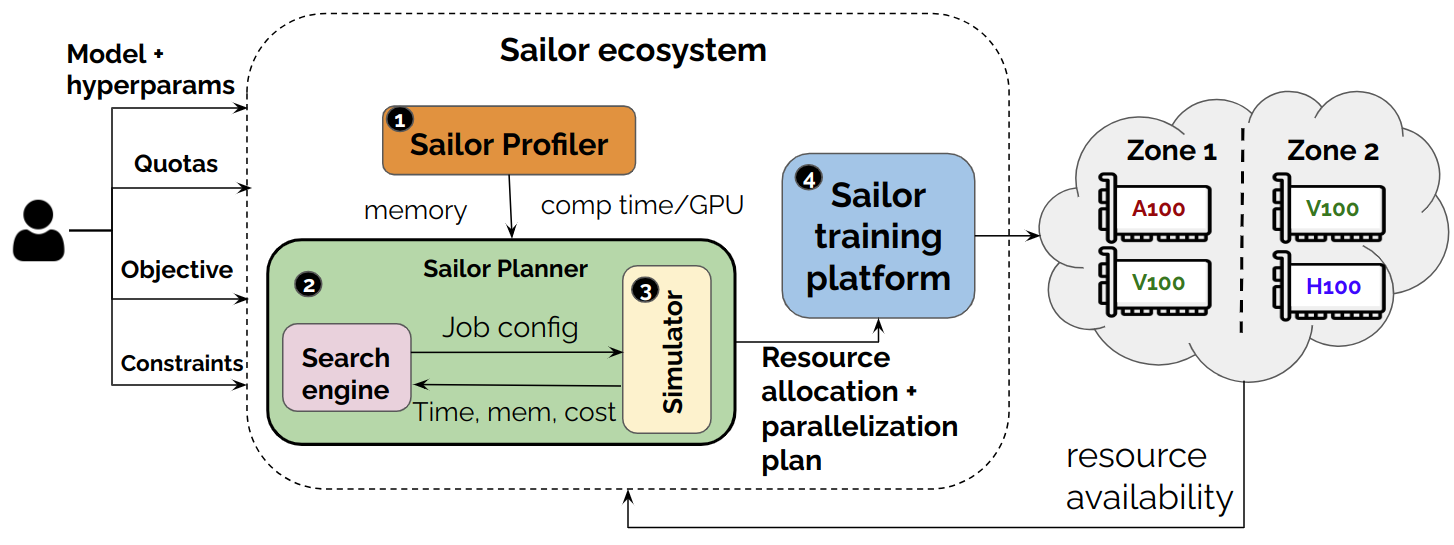}
    \caption{\sailor system overview. 
    }
\label{fig:sailor_system}
\end{figure}

To address the above challenges, we propose \textit{\sailor}, a distributed training ecosystem that consists of a profiler, planner, simulator, and distributed training framework.  
As shown in Figure~\ref{fig:sailor_system}, ML developers submit their model training specifications (model, optimizer, global batch size, etc), resource quotas (the maximum number of GPUs for each type and zone), an objective (e.g., maximize throughput or minimize cost), and optionally also constraints (e.g, a maximum budget per iteration or a maximum iteration time). \sailor also receives feedback about the current availability of hardware resources (which may be less than the quotas).

\noindent\textbf{Workflow}. 
The \sailor profiler \circled{1}  collects information about the training job, the compute nodes and network bandwidth (\S \ref{profiler}). The \sailor planner \circled{2} uses this information to select a near-optimal \textit{resource allocation} from the pool of available hardware and a \textit{job parallelization plan} that optimizes the user's objective within the provided constraints (\S\ref{planner}). The planner uses the simulator \circled{3} to accurately evaluate various candidate plans (\S\ref{simulator}) in terms of throughput, memory footprint, and cost. \sailor then launches the job with the selected configuration using its distributed training framework \circled{4} (\S\ref{framework}), which is implemented on top of Megatron-DeepSpeed~\cite{megatron-deepspeed}. \sailor dynamically re-configures the job as resource availability changes.

\subsection{\sailor Profiler}\label{profiler}

\textbf{Training job profiling:} When a user submits a training job, the \sailor profiler collects information about the job's compute and memory requirements. \sailor profiles a training job on a \textit{single} GPU node for each different GPU node type in the available resource pool. To minimize profiling time and enable single-node profiling, the profiler reduces repeated layers to a single instance (e.g., it uses one transformer layer for a given LLM).  We use PyTorch hooks~\cite{torch_hooks} to collect information per layer: the time required for the forward pass, backward pass, and update phase with different microbatch sizes and tensor parallel degrees. We use CUDA Events for accurate GPU measurements~\cite{torch_cuda_events}. We also track the number of parameters, output activation, and the memory required for intermediate stages per layer, using the PyTorch CUDA memory allocator~\cite{torch_cuda_mem_alloc}. The profiling overhead is negligible (a couple of minutes) \new{for the LLMs we consider. For models with non-uniform layers, all layers need to be profiled, which increases profiling time. Our profiling approach can be used with any dense layers, while profiling Mixture-of-Experts (where layer load may vary dynamically during training) is left for future work.} 

\textbf{Cluster profiling:} \sailor also collects information about the network bandwidth between any pair of different machine types. Since network bandwidth depends on the message size, \sailor collects network bandwidth measurements (using PyTorch collectives with NCCL backend) by varying the message size and fitting a polynomial function to get a set of coefficients for any pair of node types.
Our profiling methodology applies to all GPU types.
Adding a new GPU type requires collecting model and cluster profiling data as described above.

\subsection{\sailor Planner}\label{planner}

The \sailor planner takes as input the training job and cluster profiling information from the profiler, a performance or cost objective, and optionally a constraint such as a budget. The planner selects a resource allocation and a parallelization plan to optimize the objective under the constraints.
The parallelization plan defines the number of pipeline stages $P$ (which we refer to simply as \textit{stages}), the data parallelism degree of each stage $D$, the $D$ pairs of $(GPU_j, TP_j, Zone_j)$ for each stage, and the microbatch size $mbs$. The \sailor planner does \underline{not} change the global batch size, thus it does not affect the job's training dynamics. 

The combination of resource allocation and parallelization plan candidates creates a vast search space. To find efficient solutions \textit{quickly}, \sailor: 1) prunes the search space with heuristics that consider training memory footprint, GPU capacity, and scalability constraints (\S\ref{planner_heuristics}), and 2) applies dynamic programming to reuse information about the performance of parallelization plans (\S\ref{planner_dp}). These techniques allow \sailor to find training plans in 10s of seconds even for large heterogeneous, geo-distributed scenarios (\S\ref{sec:eval_planner}). 

The planner iterates through different parallelism degrees and microbatch sizes (based on model characteristics and profiling information). For a given layer partitioning and microbatch size, it finds the tensor parallelism degree for each GPU type, based on memory constraints and scaling heuristics. Then, it iterates through all cloud region combinations and selects the data parallelism degrees to evaluate for a combination. For a fixed data parallelism degree, the planner invokes the $solve\_dp()$ function (Listing \ref{alg:dp}) that applies dynamic programming to determine the optimal stage configuration (\S \ref{planner_dp}). The planner considers the configuration valid only when it is within the user-specified constraint.

Finally, the planner sorts the configurations according to the objective and returns the best configuration.

\subsubsection{Pruning the large search space with heuristics}\label{planner_heuristics} We introduce heuristics to prune the search, omitting cases early-on that would lead to suboptimal or invalid results:

\textit{H1: Limit tensor parallelism within a node.}\label{H1} Tensor parallelism performance is known to degrade when spanning multiple nodes~\cite{shoeybi2020megatronlm, strati2024dejavu}. We restrict tensor parallelism to a single node and do not explore cross-node pairs (unlike Metis~\cite{Um24Metis}). As a result, each tensor parallel replica of a stage only uses a single GPU type.

\textit{H2: Prune OOM configurations early.}  Since each stage replica performs tensor parallelism only among GPUs of the same type, we can easily compute the \textit{minimum} tensor parallelism degree of each GPU type, for a given pipeline parallel stage and microbatch size. We exclude cases with tensor parallelism below this minimum. To find the minimum tensor parallelism degree per GPU for a stage, we compute the memory footprint of that stage as described in \S\ref{simulator}, and identify the minimum number of GPUs required based on the available memory per GPU. The minimum tensor parallelism for each stage is independent of the number of available GPUs per type, so we can reuse it when resource availability changes.

\textit{H3: When maximizing throughput, consider data parallelism degrees in decreasing order, until throughput stops increasing.} \sailor uses the same data parallelism for each stage.  We observe that, with a fixed pipeline parallel degree, increasing data parallelism (by using more machines) benefits training throughput as more pipelines process minibatches independently. However, as the data parallelism degree increases, the time required for gradient synchronization also increases, negatively affecting training throughput. Thus, when optimizing for throughput, we first determine the maximum feasible data parallelism degree based on available resources, and then progressively reduce it, until throughput stops improving.

\textit{H4: When minimizing cost, consider data parallelism degrees in increasing order, until cost per iteration stops decreasing.} Following the logic from \textit{H3}, for a fixed pipeline, doubling the data parallelism degree will lead to doubling the number of resources, but will not half the iteration time (due to all-reduce scaling overheads). Thus, configurations with a lower data parallelism degree lead to lower cost/iteration. Thus, when the user objective is minimizing cost/iteration, we search for \textit{increasing} data parallel degrees $D$ until a solution within the throughput constraint is found.

\textit{H5: Keep data parallel communication within a single region, while spreading pipeline parallel communication across more than one region.} As shown by earlier work~\cite{strati24crossregion, palak2024atlas}, data parallelism performs poorly across regions due to the low network bandwidth. We constraint all data parallel pairs of a stage within a single region.

\textit{H6: Treat multiple zones within the same region as a single zone.} Within a cloud region, the network bandwidth across zones is  similar to the network bandwidth within a zone~\cite{strati24crossregion}. Thus, to reduce the search space, we consolidate all zones in a region into a single zone and do the geo-distributed partitioning at a region granularity.

\subsubsection{Selecting per-stage configurations with dynamic programming}\label{planner_dp}
We now describe how we find optimal resource configuration per stage for a given pipeline and data parallelism degree, microbatch size, tensor parallel degrees per stage, and GPU type. The formulation we describe below assumes an iteration time minimization objective and \S\ref{planner_dp_cost} describes how we further incorporate cost constraints. For brevity, we omit the reverse optimization (monetary cost minimization under throughput constraints).

\textbf{Problem formulation and goal:} Given a pipeline parallel degree $P$, a microbatch size $mbs$, a data parallel degree $D$, and tensor parallel degrees $tp_{ij}$ for GPU $j$ and stage $i$, we want to find, for each stage $i$,  the $D$ replicas, where each replica is a tuple $(j, tp_{ij}, zone_k)$ that minimizes iteration time. Note that \sailor precomputes $tp_{ij}$ (Heuristic \textit{H2}). 

\textbf{Why dynamic programming?} Optimizing resource allocation per stage is challenging, as assigning resources to one stage impacts overall runtime and availability for other stages. In a heterogeneous, multi-zone setup, the number of possible resource combinations per stage explodes. Related works on homogeneous clusters used  Integer Linear Programming (ILP)~\cite{zheng2022alpa, Jang2023Oobleck}, exhaustive search ~\cite{Athlur22Varuna}, or dynamic programming~\cite{li2022amp, Tarnawski21Piper}. We adopt dynamic programming for its ability to decompose the problem into subproblems and reuse intermediate results.

\textbf{Dynamic programming formulation:} Assume we have a pipeline with degree $P$, and we want to solve the dynamic programming problem for stage $i$ that has $L$ layers, with $l0$ being the first layer. We formulate selecting a resource configuration per stage as finding $r$ resources to give to stage $i$ while minimizing the iteration time $T_{iter}$ as follows: \begin{equation}\label{eq:dp}
T_{iter}[l_0][P][R] = min_r\{T_{total}(T_{iter}[l_0+L][P-1][R-r], T_i(r))\}
\end{equation}
for a given pipeline parallel degree $P$, stage $i$, and available resources $R$. 
We give $r$ resources to stage $i$, and $R-r$ resources are available for the subsequent stages. In \sailor, $r$ represents a map of different GPU types and zones, and should be enough to get $D$ replicas of this stage. Based on heuristic \textit{H5}, we keep each stage within a single region. $T_{total}$ is calculated by identifying the straggler between stage $i$ and the rest pipeline, the pipeline communication cost between stage $i$ and $i+1$, and the synchronization bottleneck between stage $i$ and the rest pipeline.

More specifically, assuming $X-i$ is the pipeline without stage $i$, and $N_b$ is the number of microbatches processed per pipeline, $t_j$ is the time per stage, and $sync_j$ is the time for synchronization:
$$
T_{total} = max(t_i(r), Straggler_{X-i}(R-r)) \cdot N_b \newline
$$
$$
+ max(sync_i(r), sync_{X-i}(R-r)) + \sum_0^{P-1} t_j 
$$
Our formulation aims to minimize iteration time and reduce pipeline stragglers, inherently excluding imbalanced pipelines with large stage time differences.

\textbf{Implementation:} Listing \ref{alg:dp} implements the above formulation for heterogeneous GPUs and various cloud zones and regions. The procedure starts by generating all resource combinations of different GPU types that could fit this stage with the specified data parallelism (line 2). Then, for each combination $r$, it finds the next available region that fits stage $i$ with $r$ resources (lines 6 and 14). If $i$ is the last stage, it returns the configuration that minimizes the stage time.

\subsubsection{Adding a monetary cost constraint}\label{planner_dp_cost}

So far, we have formulated the iteration time minimization as a dynamic programming problem, and we split the problem at per-stage subproblems as shown in Eq~\ref{eq:dp} and Listing \ref{alg:dp}. When introducing a budget constraint, we need to account for the \textit{remaining} budget per stage, to solve the DP subproblem for that stage. The monetary cost depends both on the resources used and the iteration time. However, the iteration time depends on the pipeline straggler, which is not yet known when solving the resource allocation subproblem for stage $i$. To overcome this limitation, when solving the subproblem for stage $i$, we approximate the remaining budget by first considering that stage $i$ is the straggler, and solving for the remaining stages with the respective remaining budget. At the end, we determine the actual straggler. If our straggler assumption was not correct,  we adjust the budget with the new straggler and we solve the subproblems again.

\textbf{Problem formulation:} The monetary cost per iteration includes the cost due to compute resources $C_{comp}$ and due to data transfer $C_{comm}$. When introducing a monetary cost constraint $C$, the cost per iteration should satisfy:
$$
C_{iter} = C_{comp} + C_{comm} <= C
=>\sum_{i=0}^{i=P-1}{Ccomp_i}\cdot T_{iter} + C_{comm} <= C  \newline
$$
$$
\sum_{0}^{P-1}{Ccomp_i}\cdot (\sum_0^{P-1}{t_i} + N_b \cdot t_{straggler} + max_0^P(t_{sync_i}) ) + C_{comm} <= C 
$$ where $\sum_0^{P-1}{t_i}$ stands for the pipeline warmup and cooldown phase, $N_b$ is the number of microbatches processed per pipeline, and $t_{sync_i}$ is the time required for the synchronization of all replicas of stage $i$. Since large models usually train with large global batch sizes, the $N_b \cdot t_{straggler}$ term usually determines the iteration time, we can rewrite as:
\begin{equation}\label{eq:cost_simple}
\sum_{0}^{P-1}{C_{{comp}_i}}\cdot (N_b \cdot t_{straggler})  + C_{comm} <= C 
\end{equation}
From a dynamic programming perspective, assuming we are at pipeline stage $i$, which has a maximum budget limit $C_{cur}$, the cost constraint will be: $C_i + C_{rem} <= C_{cur}$, where $C_{rem}$ is the cost of the remaining stages, and $C_i$ is the cost of stage $i$. From Eq.~\ref{eq:cost_simple}, we have that $C_i = C_{comp_i} \cdot (N_b \cdot t_{straggler})  + C_{comm_i}
$.
In Listing \ref{alg:dp}, line 14, when exploring a resource combination $r$, we can easily find $C_{comp_i}$, and $C_{comm_i}$ for stage $i$. Since we do not know $t_{straggler}$, which is required to specify the remaining budget for the next stages, we use the approximation in lines 17-32: We begin assuming stage $i$ is the straggler ($t_{straggler}==t_i$), and compute the $C_{rem}$ for the next stages accordingly. We call the \textit{solve\_dp} function for the next stages giving $C_{rem}$ as the budget constraint (line 20). If a solution cannot be found, we proceed with the next resource combination for stage $i$ (line 22). If a solution is found,  we check whether it is within the cost limit and keep the one with the maximum throughput (line 26). We also check the straggler of the found solution: if it is the same with the one we assumed, we break, and proceed to the next resource combination (lines 26-28). Otherwise, we adjust the budget with the new straggler (lines 31-32) and iterate again.

\begin{lstlisting}[float, caption=Resource-Stage assignment algorithm, basicstyle=\small, label=alg:dp]
def solve_dp(P, i, R, D, tp_gpu_stages, Ci):
    Rcombos = generate_combos(R,D,tp_gpu_stages[i])
    if i==P-1:
        time_i = None
        for r in Rcombos:
            next_region = find_region_fits(i, r, curr_region)
            time_ir = time_for_stage(i,r,next_region)
            cost_ir = cost_for_stage(i, r, next_region, time_ir)
            if cost_ir <= Ci: time_i = min(time_i, time_ir)
        T_iter[P][i][R] = time_i
    else:
        time_all = None, min_time = None
        for r in Rcombos:
            next_region = find_region_fits(i, r, curr_region)
            time_i = time_for_stage(i,r,next_region)
            cost_i = cost_for_stage(i, r, next_region, time_i)
            C_rem = Ci - cost_i
            assumed_straggler = time_i
            while (C_rem > 0):
                nextconf = solve_dp(P, i+1, R-r,
                    next_region, all_regions, C_rem)
                if nextconf is None: break
                time_all = total_time(time_i, nextconf.time)
                cost_all = total_cost(cost_i, nextconf.cost)
                if nextconf.straggler < assumed_straggler:
                    if (cost_all <= Ci):
                        min_time = min(time_all, min_time)
                    break
                cost_i = cost_for_stage(i, r, next_region,
                    straggler=nextconf.straggler)
                C_rem = Ci - cost_i
                assumed_straggler = nextconf.straggler
        T_iter[P][i][R] = min_time
    return  T_iter[P][i][R]
\end{lstlisting} \label{alg:dp}

\subsection{\sailor Simulator}\label{simulator}

The planner uses the simulator to evaluate the performance and memory footprint of the generated plans. The \sailor simulator takes as input a training job specification (model, global batch size, optimizer, hyperparameters) and a job parallelization plan. It then estimates the memory footprint per GPU, the iteration time, and the cost per iteration. The simulator allows the planner to easily specify different types of heterogeneity: hardware heterogeneity (GPU type, number of GPUs per node, and network bandwidth) and job configuration heterogeneity (number of pipelines and different stage configurations per pipeline). 
The training configuration also specifies the microbatch size. 
The simulator also incorporates the information collected by the profiler about the training job and network bandwidth of the used links.  Both the \sailor planner and simulator treat GPUs as black-box compute units, thus they can seamlessly support GPUs from different generations, vendors, and even different accelerators (e.g. TPUs).

\textbf{Memory footprint estimation:} The \sailor simulator accurately estimates a training job's memory footprint by: 1) calculating memory footprint per GPU, per-stage and 2) considering \textit{all} main sources of memory footprint during training. Compared to prior works that assume a homogeneous memory footprint per stage, we observe that for a given parallelism configuration, the memory footprint of a training worker depends on its layer partitioning, pipeline stage index, tensor parallelism degree, and microbatch size. Hence, memory footprint varies among workers and needs to be analyzed per worker to detect OOM scenarios.

Second, the peak memory footprint of a worker throughout training consists of various sources, often ignored in prior works. The peak memory footprint $M_{peak}$ of a worker is given by: $M_{peak} = M_{model} + M_{activation}$, where $M_{model}$ corresponds to the memory needed to keep copies of model parameters and is given by $M_{model} = num\_params \cdot mul\_factor \cdot data\_type\_size$. $num\_params$ is found by the stage id and tensor parallelism degree of the worker, while $mul\_factor$ accounts for multiple copies needed for the model itself, the optimizer, gradients, and communication~\cite{Rajbhandari21Zero}. $M_{activation}$ is the memory needed for storing layer activations and depends on the stage id and tensor parallelism degree of a worker, and microbatch size. By computing the memory footprint for each worker and comparing with the worker's GPU capacity, the simulator can easily detect OOM cases.

\textbf{Iteration time estimation:} We define one iteration as a full pass over the user-defined global batch size. The iteration time is calculated as:
$T_{iter} = \max(T_{ppi}) + T_{sync} + T_{update}$, where $T_{ppi}$ is the time needed for pipeline $i$, $T_{sync}$ is the time needed for gradient synchronization at the end of an iteration, and $T_{update}$ is the time for model update. We compute $T_{{pp}_i}$ and $T_{sync}$ following the formulas described in \cite{strati24crossregion} for 1F1B pipeline parallelism,  using our profiling for network bandwidth with respect to the message size per network link, to estimate communication time (for peer-to-peer and collectives). For each pipeline, the 1F1B pipeline parallelism schedule includes a warm-up, steady, and cool-down phase per iteration, where the steady phase is determined by the stage with the largest computation time (straggler). After computing the iteration time per pipeline, we compute synchronization time. Taking the maximum time per pipeline accounts for straggler effects caused by heterogeneity in GPU generations, inter-GPU, and CPU-GPU interconnects.

\textbf{Iteration cost estimation:} Related works (Table \ref{table:planners}) do not compute the monetary cost of different resource allocation and job configuration combinations, since they only optimize for throughput. However, a very important metric to account for is \textit{cost per iteration}, especially with geo-distributed training, due to costs associated with across-zone communication. Since \sailor does not change the global batch size and training hyperparameters, the number of iterations needed to reach convergence is constant, regardless of the cluster setup and parallelization strategy. Thus, the monetary cost per iteration indicates the total budget needed for the whole training. The metric depends both on the cost of allocated resources, as well as the training throughput and communication cost. The cost per iteration is given by $C_{iter} = C_{comp} + C_{comm}$, where $C_{comp}$ is the cost due to compute resources, and is calculated as $\sum_i({N_i \cdot cost\_per\_gpu_i}) \cdot T_{iter}$, for all different GPU types $i$ in the cluster. $C_{comm}$ stands for the cost for data exchange per iteration (e.g., when using geo-distributed training in public cloud) and is given by $C_{comm} = \sum_{ij}(bytes_{ij} \cdot cost\_per\_byte\_{ij})$ for all zones $i,j$ in the cluster. $bytes_{ij}$ might include traffic for data and pipeline parallel communication.

\subsection{\sailor Distributed Training Framework}\label{framework}

The \sailor training framework receives the job configuration that the planner generates, sets up the cluster, and starts training. We modified Megatron-DeepSpeed~\cite{megatron-deepspeed} to support heterogeneous plans and seamless elasticity, to make it compatible to planner's output and allow for fast reconfiguration when resource availability changes.

\textbf{Support for heterogeneous plans:} We added support for varying tensor-parallel degrees across data-parallel pairs per pipeline stage. The framework takes as input a rank topology for each stage, allowing each rank to belong to distinct tensor-parallel groups. Different tensor parallelism per stage affects the pipeline and data parallel communication, requiring workers to split or replicate activations and gradients across multiple peers. To accommodate this, we adjust the PyTorch communication groups and modify the send/recv and all-reduce operations accordingly.

\textbf{Support for fault-tolerance and elasticity:} The Megatron-DeepSpeed framework lacks failure recovery and dynamic resource reconfiguration: the whole training needs to stop, and the user needs to \textit{manually} reconfigure and restart the job. However, resource availability frequently changes both in the cloud (especially with spot instances)~\cite{Thorpe23Bamboo, strati24crossregion} and in on-premise datacenters as jobs start or finish~\cite{wagenlander24tenplex}. We introduce modifications for fast reconfiguration. Each job consists of a controller and multiple workers. The workers handle training, while the controller monitors their status and detects resource availability changes. Upon detecting a change, the controller reinvokes the planner to generate a new plan and instructs workers to adjust accordingly. We follow a kill-free approach to minimize reconfiguration time: existing workers destroy the current communication group, clean up their GPU memory, repartition the model, and setup a new communication group.  If additional resources become available, the controller waits for new workers to initialize before updating the training configuration.  Training restarts from the latest available checkpoint.
We use asynchronous checkpointing to minimize the rollback time~\cite{Mohan21CheckFreq, strati25pccheck}.

\section{Evaluation}\label{sec:eval}

We evaluate \sailor to answer the following questions:

\begin{enumerate}
    \item How accurately does the \sailor simulator estimate iteration time and memory footprint?
    \item How well does the \sailor planner perform compared to baselines in homogeneous and heterogeneous setups, in terms of throughput and monetary cost?
     \item How is the \sailor planner search time affected by the cluster size, resource heterogeneity, user constraints and the different optimizations?

\end{enumerate}

\textbf{System configurations: } We evaluate \sailor using real hardware and simulations. We use different cluster setups and GPU generations, in both cloud environments and on-premise clusters. In the public cloud, we used VMs with A100-40GB and V100-16GB  from GCP~\cite{gcpgpus}. For our on-premise datacenter experiments, we used a cluster with up to 32 homogeneous machines with 4 Grace-Hopper GPUs each, and a cluster of heterogeneous machines with 2x8 Titan-RTX, 3x8 RTX-2080, and 2x8 RTX-3090. 

\textbf{Models:} We use the OPT-350M~\cite{hfopt} and GPT-Neo-2.7B~\cite{hfgptneo} models, with global batch size of 2048 sequences, and sequence length of 2048 tokens, with the Adam optimizer.

\textbf{Baselines: } To our knowledge, we present the first comprehensive comparison of major planners for large-scale ML training, including planners targeting \textit{homogeneous} resources (Piper~\cite{Tarnawski21Piper}, Varuna~\cite{Athlur22Varuna}, Aceso~\cite{Liu24Aceso}), \textit{heterogeneous} resources (AMP~\cite{li2022amp}, Metis~\cite{Um24Metis}, FlashFlex~\cite{yan2024flashflex}), and \textit{geo-distributed training} (DTFM\cite{yuan2023decentralized}). All baselines, except Aceso, are integrated into our platform with a unified Python API. We profile our models once and give each baseline its required profiling information.  Aceso defines its own operators, which we profile separately. As Aceso also uses the Megatron backend, per-layer runtime profiles are very close to our models. 
We used the open-source version of all baselines. DTFM~\cite{yuan2023decentralized} does not determine parallelization strategies (e.g., DP, PP), but instead partitions a given plan. Therefore, we exhaustively generated all homogeneous parallelization plans and applied their partitioning methods to each. As the Atlas paper does not have an open-source implementation of runtime and memory simulation, we were unable to test Atlas end-to-end. However, we tested the zone assignment described in the paper~\cite{palak2024atlas}, which performs similar to \sailor.

\subsection{Validation of the \sailor simulator}\label{sec:eval_val}

We evaluate the accuracy of the \sailor simulator's iteration time and memory footprint estimations. We vary the number of devices and parallelization plans, find the difference in iteration time and peak memory footprint, and summarize using box plots. We omit AMP and DTFM since they do not support memory estimation.

\begin{figure}[t!]
    \centering
    
    \begin{subfigure}[b]{\linewidth}
        \centering
        \includegraphics[width=\linewidth]{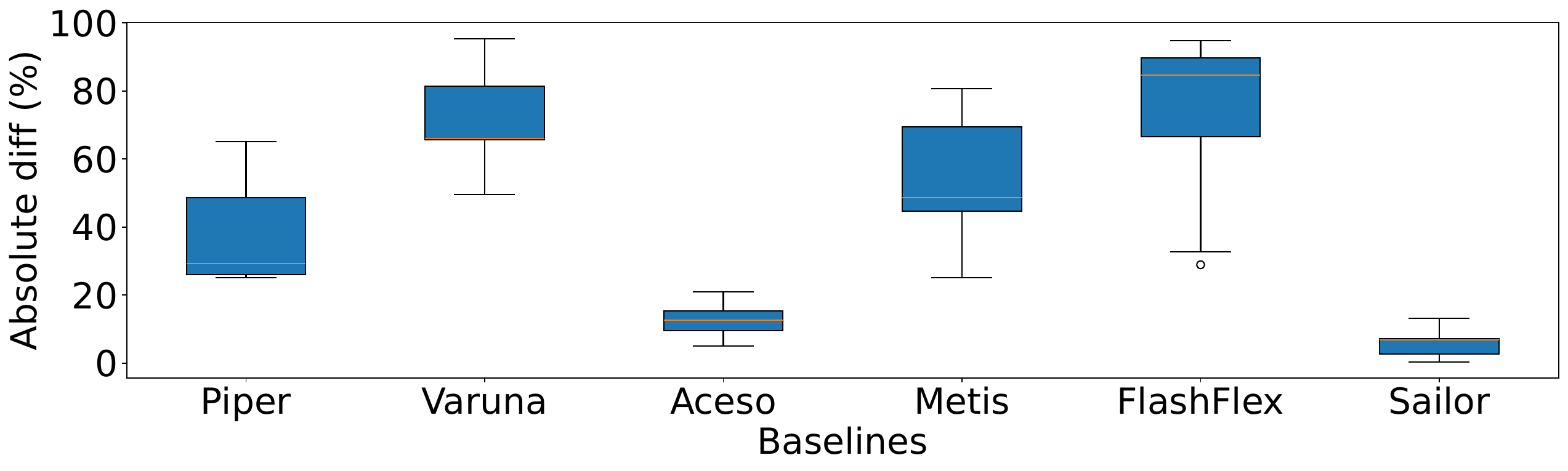}
        \caption{Peak memory footprint estimation}
        \label{fig:val_homo_mem}
    \end{subfigure}

    \begin{subfigure}[b]{\linewidth}
        \centering
        \includegraphics[width=\linewidth]{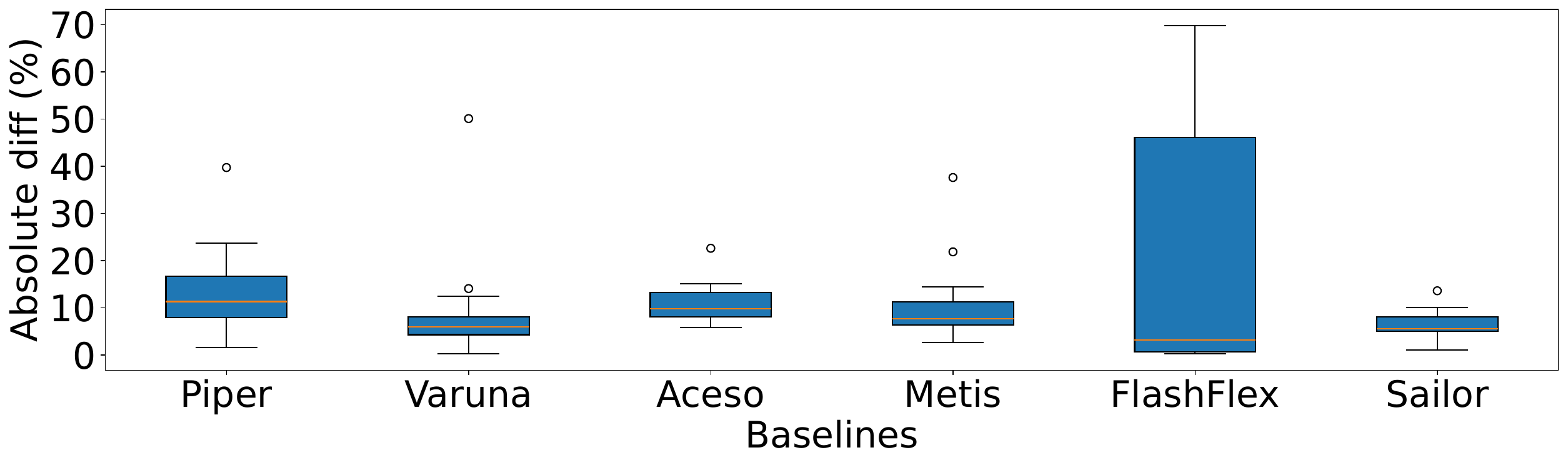}
        \caption{Iteration time estimation}
        \label{fig:val_homo_time}
    \end{subfigure}
    \caption{Different planners' compute and memory estimation on a cluster of GH200 GPU for the OPT-350M model. 
    }
    \label{fig:val_homo}
\end{figure}

\textbf{Cluster of homogeneous GPU types.} Figures \ref{fig:val_homo_mem} and \ref{fig:val_homo_time} show the iteration time and peak memory footprint estimation error for the homogeneous cluster of Grace-Hopper for the OPT-350M model, respectively. Most baselines fail to accurately capture the peak memory footprint, since they ignore significant memory sources and assume a homogeneous memory footprint across the different pipeline stages. The examined baselines exhibit an error of 12.5-74\% on average, while \sailor  achieves an average error of 5.56\%. \sailor also reduces the average runtime estimation error in the homogeneous setup to 6\%, compared to 10-20\% for the baselines.

\begin{figure}[t!]
    \centering
    \includegraphics[width=\linewidth]{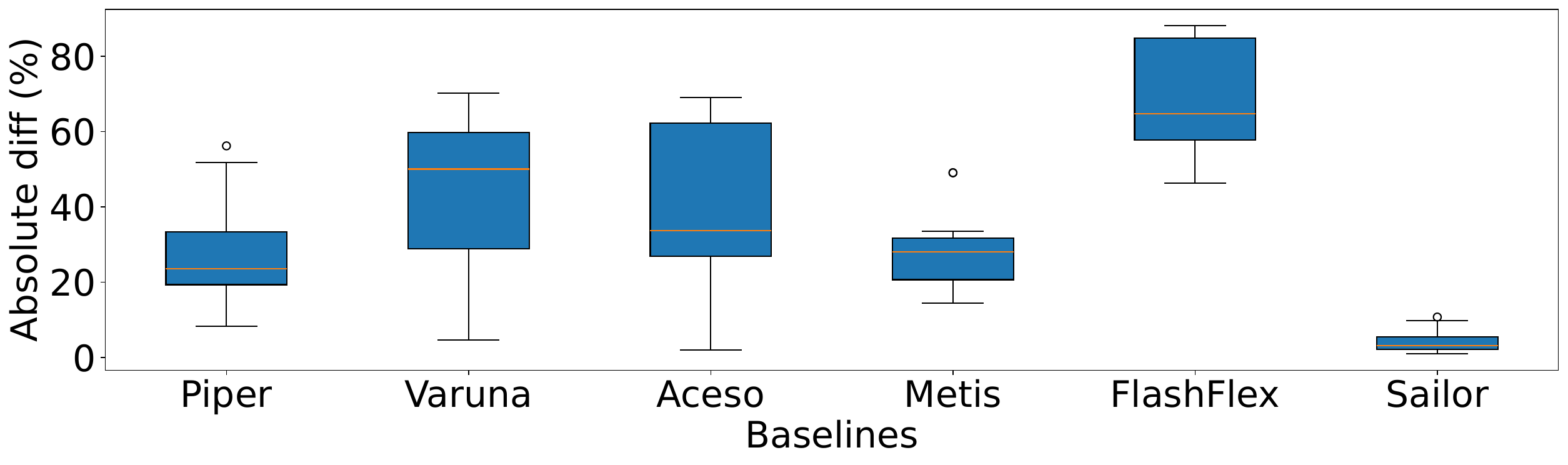}
    \caption{Different planners' iteration time on a cluster with heterogeneous GPUs for the OPT-350M model.}
    \label{fig:val_time_hetero}
    \vspace{-0.4cm}
\end{figure}

\textbf{Cluster of Heterogeneous GPU types.} Figure \ref{fig:val_time_hetero} shows the iteration time prediction error for the OPT-350M model in a heterogeneous cluster of Titan-RTX, RTX-2080, and RTX-3090. The homogeneous planners (Piper, Varuna, Aceso) do not consider the differences in forward and backward passes of the different GPU types, resulting in an average error of 28\%, 47\%, and 37\%, respectively. Even heterogeneous planners fail to accurately capture runtime: since FlashFlex relies on the theoretical performance of GPUs, it cannot accurately predict the runtime, getting an error of 69\%. Metis fails to fully capture the heterogeneous network bandwidth between nodes, thus miscalculating the communication cost, resulting in 28\% error in iteration time estimation, on average.On average, \sailor's iteration time estimations error is 4.5\%.

\subsection{\sailor Planner vs. Baselines}\label{sec:eval_planner}

We evaluate the throughput achieved by \sailor and the baselines across various cluster configurations using both real hardware and our simulator. All baselines require a predefined resource topology as input: we consider 4-GPU VMs for each GPU type.
\sailor takes resource quotas as input (total \textit{number} of GPUs per type per zone) and jointly determines both the topology (VM allocation) and the parallelization plan. For Metis, we impose a 300-second time limit and use the best solution found within that period, if available. We summarize key takeaways. 

\begin{figure}[t]
    \centering
\includegraphics[width=\linewidth]{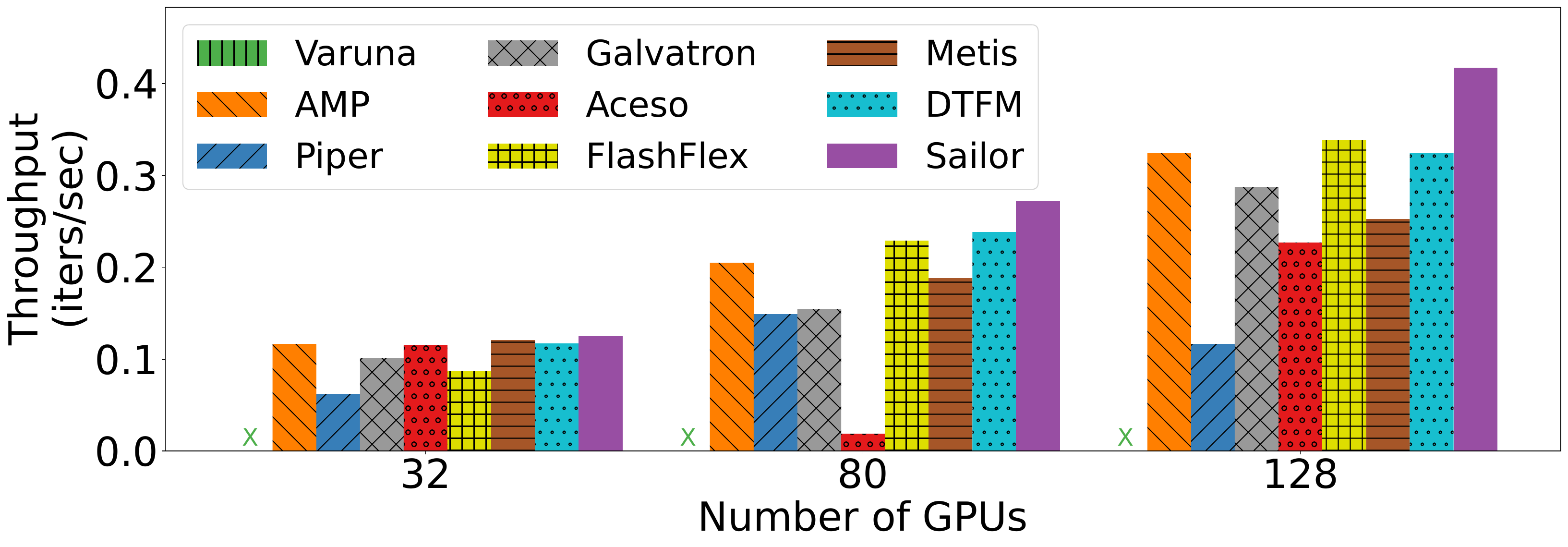}
    \caption{Comparison of the different planners considering A100-40GB GPUs for the OPT-350M model in one zone }
    \label{fig:planner_homo}
\vspace{-0.5cm}
\end{figure}

\subsubsection{Homogeneous Setups}\label{sec:eval_planner_homo}

Figure \ref{fig:planner_homo} shows the throughput achieved using the baseline planners and \sailor with only A100 GPUs for the OPT-350M model. Varuna failed to generate a valid plan that would not lead to OOM errors due to the poor memory estimation, and its limited search space (only supporting 2D parallelism). \sailor improves throughput by 1.15$\times$ compared to the closest baseline (DTFM), and even up to 5.7$\times$ (compared to Aceso).

\begin{figure*}[t!]
    \centering
    \begin{subfigure}[b]{0.49\linewidth}
        \centering
    \includegraphics[width=\linewidth]{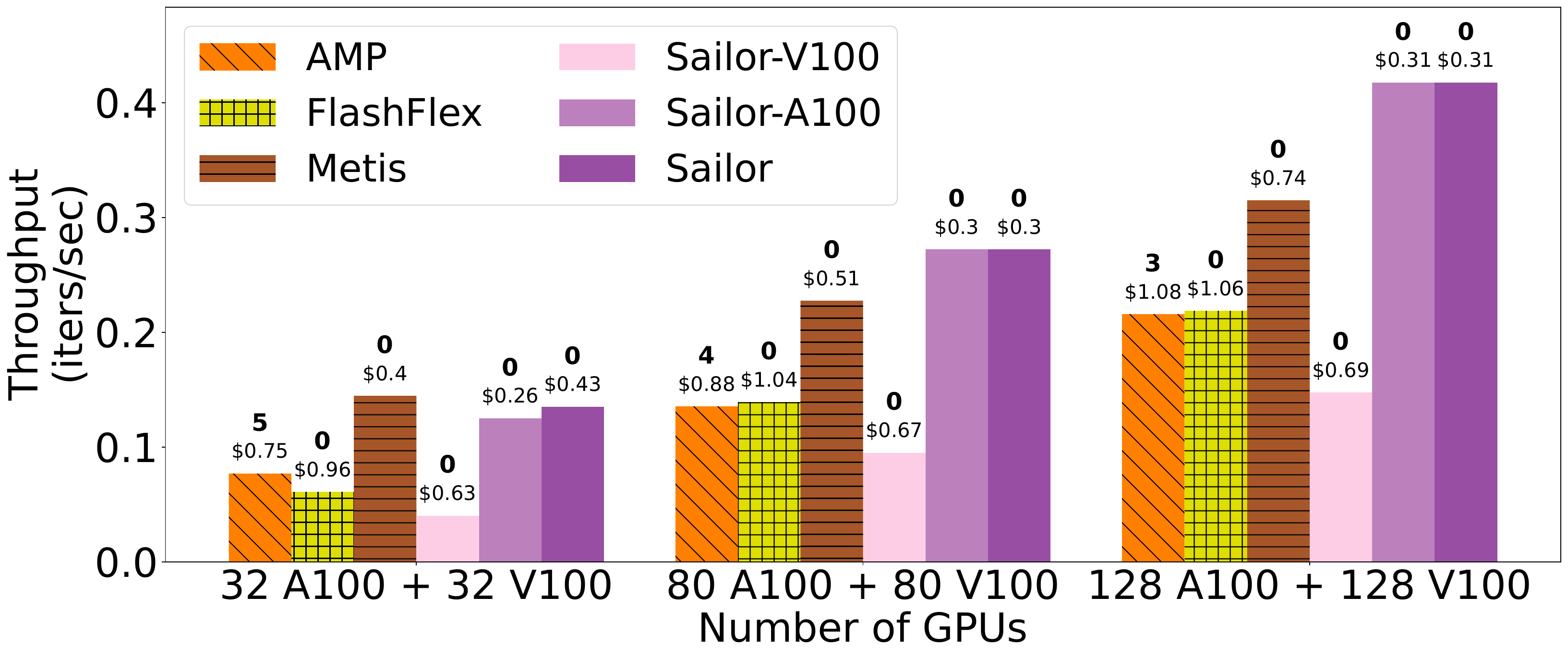}
        \caption{50\% A100, 50\% V100}
        \label{fig:planner_het_OPT_50}
    \end{subfigure}
    {\hskip 0.05in}
    \begin{subfigure}[b]{0.49\linewidth}
        \centering
        \includegraphics[width=\linewidth]{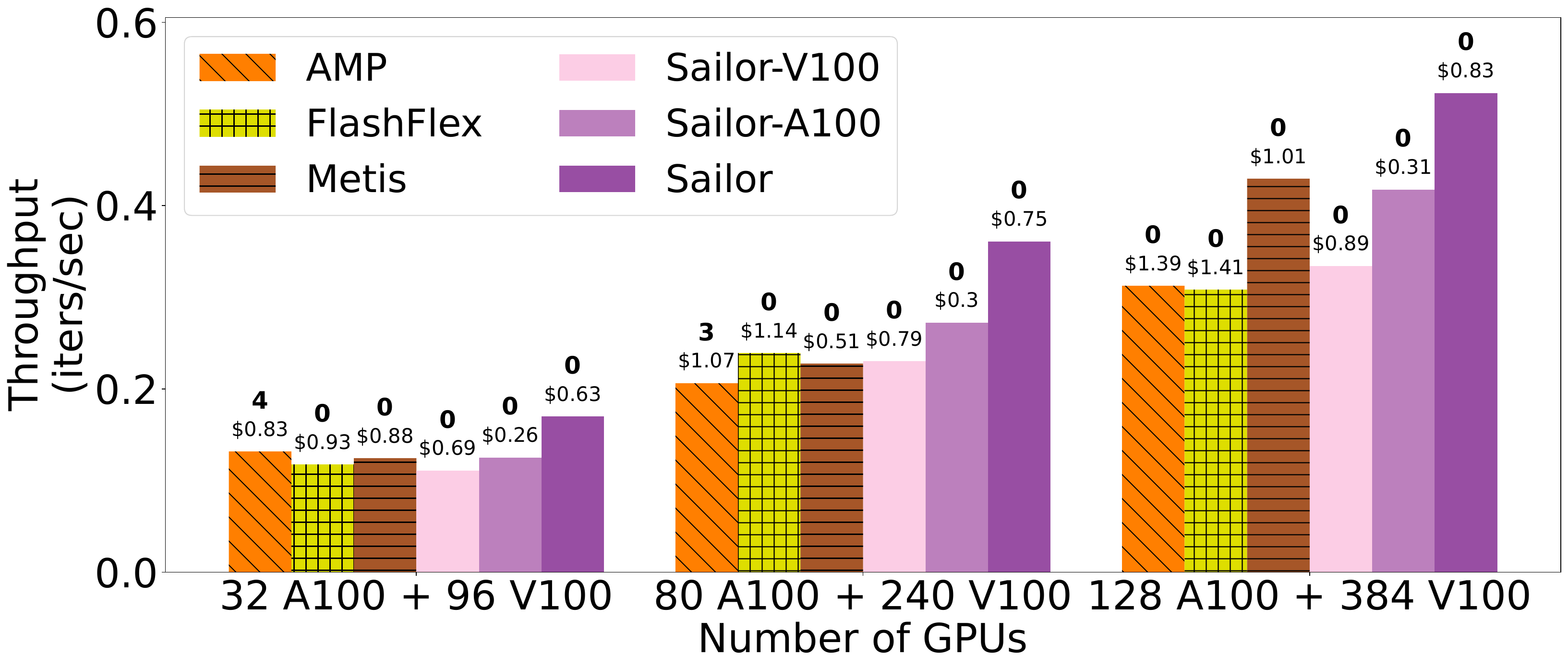}
        \caption{25\% A100, 75\% V100}
        \label{fig:planner_het_OPT_25}
    \end{subfigure}
    \caption{Comparison of planners considering A100 and V100 GPUs for the OPT-350M model in 1 zone }
    \label{fig:planner_het_OPT}
\end{figure*}

\begin{figure*}[t!]
    \centering
    \begin{subfigure}[b]{0.49\linewidth}
        \centering
    \includegraphics[width=\linewidth]{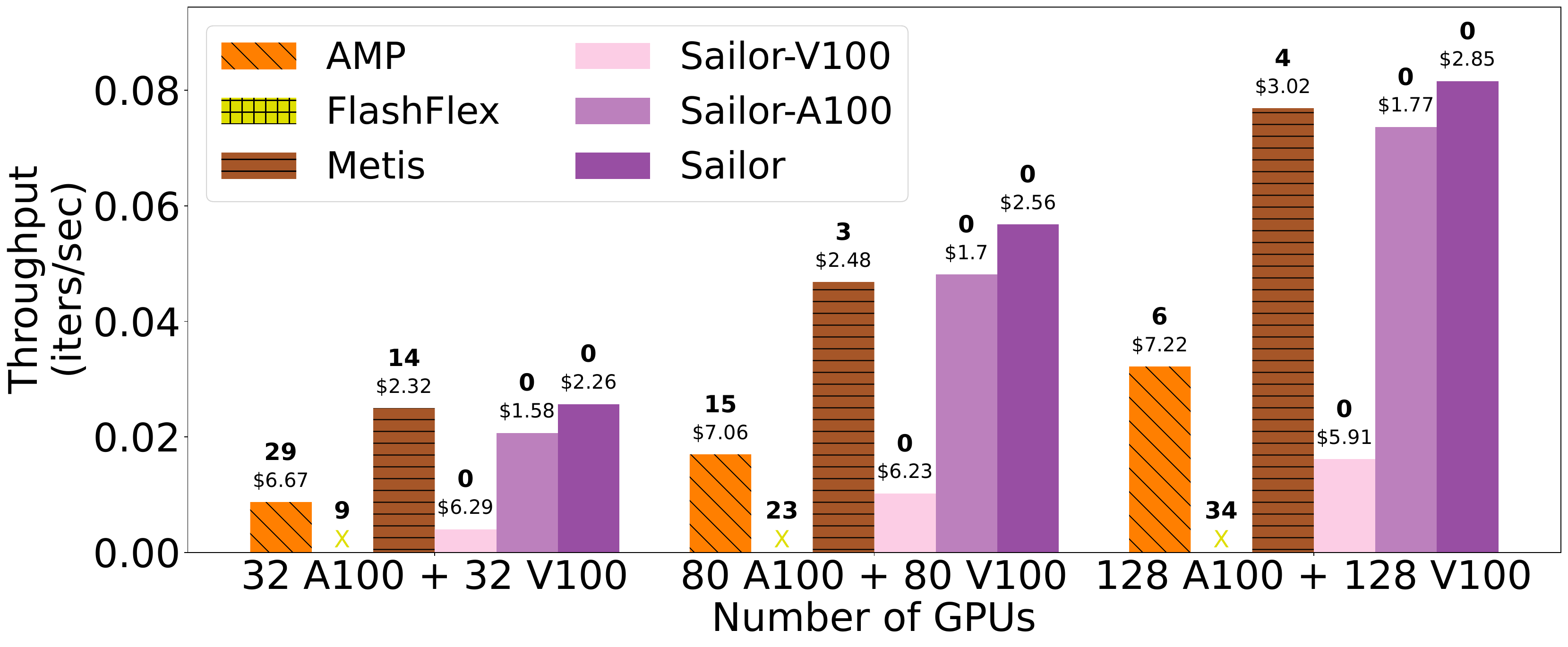}
        \caption{50\% A100, 50\% V100}
        \label{fig:planner_het_GPT_50}
    \end{subfigure}
    {\hskip 0.05in}
    \begin{subfigure}[b]{0.49\linewidth}
        \centering
        \includegraphics[width=\linewidth]{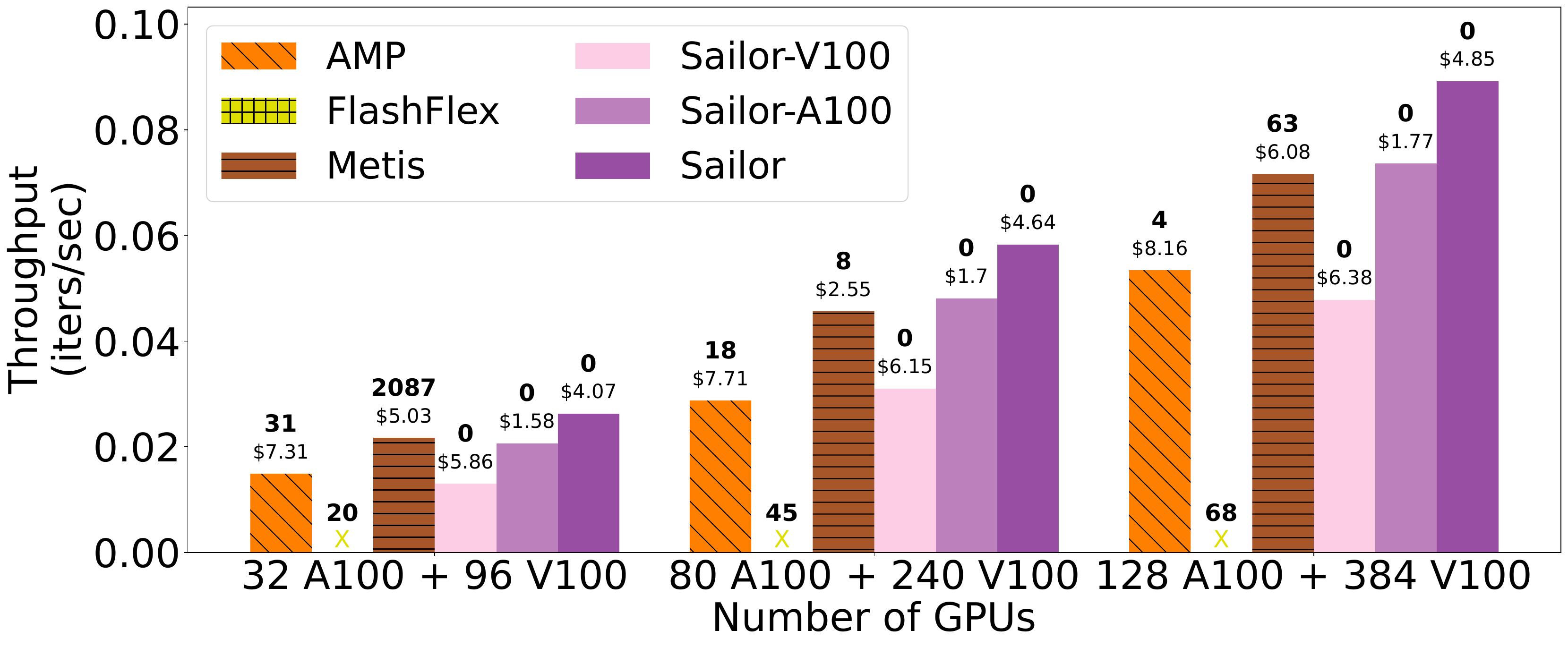}
        \caption{25\% A100, 75\% V100}
        \label{fig:planner_het_GPT_25}
    \end{subfigure}
    \caption{Comparison of planners considering A100 and V100 GPUs for the GPT-Neo-2.7B model in 1 zone }
    \label{fig:planner_het_GPT}
\end{figure*}

\subsubsection{Heterogeneous Setups}\label{sec:eval_planner_het}

We evaluate \sailor's throughput compared to baselines (AMP, Metis, and FlashFlex) in a heterogeneous cluster setup using a mixture of A100 and V100. We vary the ratio and amount of A100 and V100 (50\%-50\% and 25\%-75\%) to assess the impact of having more low-end GPUs in the cluster~\cite{guo2024cephalo}). As in the homogeneous setup, all baselines get a fixed resource topology (4-GPU VMs) as input, while \sailor also determines the resource topology along with the parallelization plan.

\textbf{Throughput of heterogeneous baselines:} Figures~\ref{fig:planner_het_OPT}~and~\ref{fig:planner_het_GPT} use our simulator to evaluate \sailor and the baselines in large heterogeneous clusters for the OPT-350M and GPT-Neo-2.7B model, respectively. We also compare the throughput achieved by \sailor, when using only homogeneous resources (either A100 or V100).  We also report the monetary cost per iteration at each scenario (number on top of bar), and the number of plans generated by the baseline that would lead to OOM before a valid plan was found (\textbf{bold} number on top of bar). Although AMP achieves high throughput in the homogeneous case,  it performs poorly in the heterogeneous scenario, as it only allows for homogeneous plans, and does not correctly model stragglers. As a result, the monetary cost per iteration also increases. Furthermore, since AMP does not model training memory footprint, it leads to a large number of OOM plans, especially in the case of the GPT-Neo model (Figure ~\ref{fig:planner_het_GPT}). 

FlashFlex achieves similar or higher throughput than AMP, as it can consider heterogeneous plans, and captures differences in compute between the different GPUs. However, its throughput is still low, as it uses low tensor parallelism and microbatch sizes. This leads to higher iteration costs (e.g. Figure \ref{fig:planner_het_OPT}), since it uses a large number of resources with a low throughput. It also fails to find valid plans for the large GPT-Neo model due to suboptimal memory estimation. Metis capped at 300 seconds achieves higher throughput than FlashFlex and AMP, due to more accurate runtime and memory footprint estimation, layer partitioning, load balancing, and exhaustive search of different GPU group combinations, but it generates a huge number of OOM plans (Figure \ref{fig:planner_het_GPT}).

\sailor achieves significantly higher throughput compared to baselines: for the OPT-350M model, \sailor achieves 1.9$\times$, 2.03$\times$, and 1.15$\times$ higher throughput compared to AMP, FlashFlex, and Metis, respectively, when the ratio of A100 and V100 GPUs is equal, and 1.57$\times$, 1.55$\times$, 1.39$\times$ higher throughput when more V100 are available. 
The speedups are similar for the GPT-Neo model as well. Compared to baselines, \sailor uses larger tensor parallelism and longer pipelines, accounting for data parallelism limitations among heterogeneous nodes (for example AMP uses data parallelism of 256 in Figure~\ref{fig:planner_geo} which significantly increases the data parallelism communication cost.) Also, \sailor does not output invalid plans, significantly improving the plan deployment compared to baselines with 10s of invalid plans. 
\sailor's ability to discover efficient resource topologies and parallelization plans also translates to significantly lower cost compared to the baselines (up to 2.67$\times$ lower cost compared to baselines in Figures \ref{fig:planner_het_OPT} and \ref{fig:planner_het_GPT}).

We also evaluated \sailor and the other heterogeneous baselines with real hardware using a smaller cluster of A100 and V100 GPUs. Figure \ref{fig:planner_het_real} shows the throughput achieved for the OPT-350M model when using an equal amount of GPUs per type (8 each), and when using more V100 than A100, as V100 were more readily available. When using the same number of GPUs per type, \sailor outperforms the baselines by 1.08-1.81$\times$ and does not generate any Out-Of-Memory plans. In contrast, AMP and Metis generate 5 and 1 invalid (OOM) plans before finding a valid plan, respectively. The valid plan found by Metis is similar to Sailor’s, but Sailor avoids invalid outputs entirely, and finds a solution in less than 1 sec, while Metis takes 60 sec.
When using 8 A100 and 16 V100 GPUs, Metis fails to output a plan as it requires the global batch size to be equally divisible by the total number of GPUs. We therefore reuse the plan from the 16 GPU case. AMP produces the same plan for the 24 GPU case as the 16 GPU case, while the plan provided by FlashFlex utilizes all 24 GPUs but uses an unnecessarily large number of pipeline stages that degrades throughput. \sailor outperforms the baselines by 1.19-2$\times$ for the 24 GPU scenario. Our simulator's estimated iteration times were within 4\% of those measured on real hardware.

\textbf{Search times:} Table~\ref{table:search_times_het} shows the search time for Figure~\ref{fig:planner_het_GPT_25} for the different baselines. Metis has long search times. In our experiments, we let Metis search for up to 300 seconds and take the best plan it produces in this time window. The other baselines are significantly faster, finishing their search in less than 200 seconds in all cases. \sailor keeps the search time within 1 minute even for the largest case (512 GPUs) due to its efficient search algorithm.

\textbf{Benefits of heterogeneity:} 
Figures \ref{fig:planner_het_OPT}, \ref{fig:planner_het_GPT} and \ref{fig:planner_het_real} also show the throughput achieved by \sailor when given homogeneous-only setups of A100 (\sailor-A100) or V100 GPUs (\sailor-V100). Given that V100s are less efficient than A100s, using A100 only, or a mixture of A100 and V100 is always better than using V100 only. However, using V100 \textit{in addition} to A100 does not always improve throughput. Heterogeneity is more beneficial when resources are limited (e.g. Figure \ref{fig:planner_het_OPT_50}, with 32 GPUs per type), or with larger models like GPT-Neo. In fact, for the OPT-350M model, when 128 A100 and 128 V100 are available, \sailor chooses a plan with 128 A100 only, as it determines that no additional benefits will be gained by adding extra resources. Moreover, when the ratio of V100 to A100 is higher than 1, the throughput improvents are more significant, as shown in Figures \ref{fig:planner_het_real}, \ref{fig:planner_het_OPT_25} and \ref{fig:planner_het_GPT_25}, where the V100:A100 ratio is 2:1 or 3:1. This aligns with the GPU memory capacity ratio, as well as the time for forward/backward pass for the transformer layers of the two GPU types, enabling better load balancing. Finally, heterogeneity leads to a higher cost, as more resources are used. Since the objective is to maximize throughput, \sailor ignored monetary cost when searching job configurations. In \S\ref{sec:eval_constraints}, we will show \sailor's ability to consider budget optimization and constraints.

\textbf{\textit{Key takeaway 1:} Heterogeneity is most beneficial when resources are limited, or for larger models, or when the ratio of the different GPU types aligns with their memory and compute characteristics for better load balancing.}

\begin{table}[t!]
\centering
\begin{tabular}{|c|c|c|c|}
\hline
\textbf{Baseline} & \multicolumn{3}{c|}{\textbf{A100-V100}} \\
\hline
& \textbf{32-96} & \textbf{80-240} & \textbf{128-384} \\
\hline
\textbf{AMP} & 31.22 & 51.43 & 86.41 \\
\hline
\textbf{FlashFlex} & 4.05 & 54.67 & 222.64 \\
\hline
\textbf{Metis} & 300 & 300 & 300\\
\hline
\textbf{\sailor} & 1.6 & 7.67 & 17.4 \\ 
\hline
\end{tabular}
\caption{Search times (in seconds) for Figure \ref{fig:planner_het_GPT_25}. }
\vspace{-0.7cm}
\label{table:search_times_het}
\end{table}

\begin{figure}[t]
    \centering
    \includegraphics[width=\linewidth]{figures/evaluation/planner/heterogeneous/fig_het_real.png}
    \caption{Throughput of heterogeneous planners for clusters of A100 and V100 GPUs for the OPT-350M model.}
    \label{fig:planner_het_real}
\vspace{-0.5cm}
\end{figure}

\subsubsection{Geo-distributed setups}\label{sec:eval_planner_geo}

\begin{figure}[t]
    \centering
    \includegraphics[width=\linewidth]{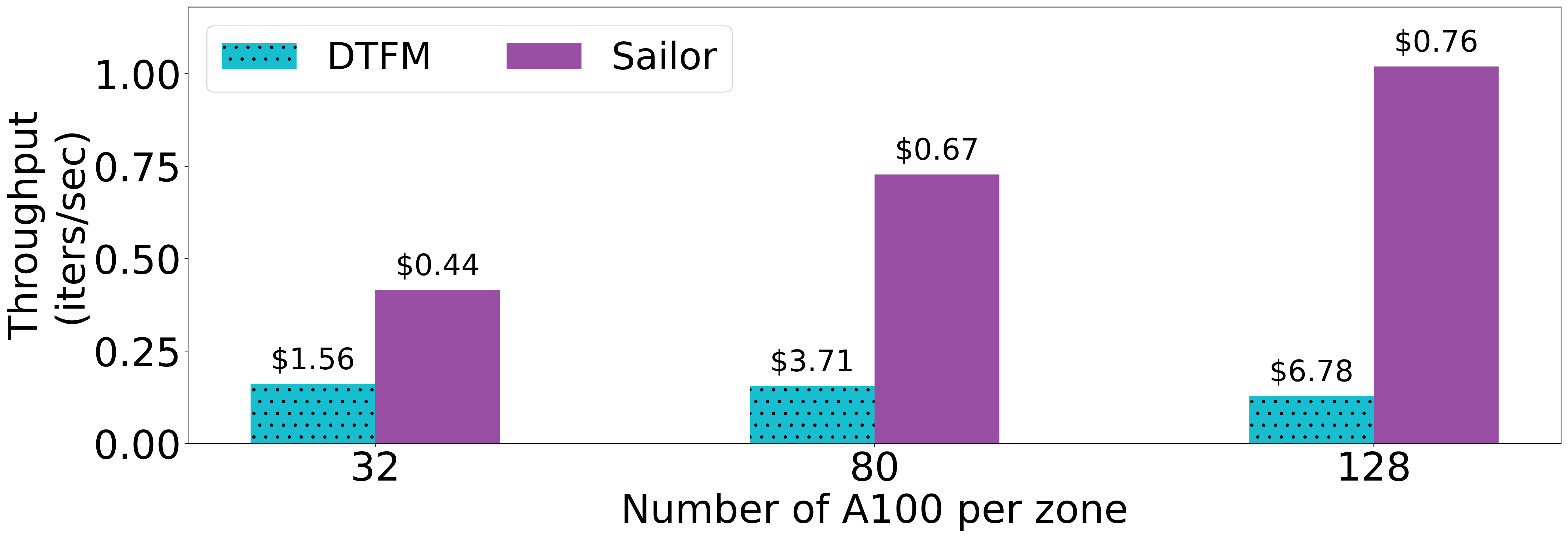}
    \caption{Throughput of geo-distributed planners with A100-40GB for the OPT-350M model in 5 zones (2 regions) using our simulator.}
    \label{fig:planner_geo}
\vspace{-0.5cm}
\end{figure}

In Figures \ref{fig:planner_geo_real} and \ref{fig:planner_geo}, we evaluate \sailor's throughput in geo-distributed setups, considering A100 GPUs for the OPT-350M model. We compare \sailor with DTFM, with the exhaustive search to automatically discover parallelization plans. We report training throughput and monetary cost per iteration, taking both the computation and communication cost into account.

\textbf{Small-scale results on real hardware:} Figure \ref{fig:planner_geo_real} shows an experiment with small-scale cluster in 4 cloud zones (2 regions) using 4 and 8 A100 GPUs per zone. DTFM cannot fully leverage multiple zones, mainly due to its suboptimal cost function and lack of memory footprint estimation. DTFM ranks solutions based on the time spent in data and pipeline parallel communication, which leads to suboptimal solution ranking. Furthermore, it uses all cloud regions, which increases communication bottlenecks and cost without increasing throughput. In contrast, \sailor uses only 1 region with all available zones (us-central1), as incorporating an additional region (us-west1) does not improve throughput. \sailor leads to 1.9$\times$ and 2.45$\times$ higher throughput than DTFM for two cluster sizes. Our simulator was within 3.7\% of the real throughput in this scenario.

\begin{figure}[t]
    \centering
    \includegraphics[width=\linewidth]{figures/evaluation/planner/geodistributed/fig_geo_real.png}
    \caption{Throughput of geo-distributed planners with A100-40GB for the OPT-350M model in 4 zones (2 regions) using real GPUs.}
    \label{fig:planner_geo_real}
\end{figure}

\textbf{Large-scale results on simulation:} In Figure \ref{fig:planner_geo} we use our simulator to evaluate larger cluster sizes with 5 cloud zones (2 regions). \sailor achieves 5.9$\times$ higher throughput and 9.48$\times$ lower cost per iteration than DTFM. \sailor employs larger microbatch sizes and tensor parallelism degrees, reducing the pipeline and data parallel data transfers.  Furthermore, \sailor finds configurations within 1 second, compared to DTFM that needs hundrend of seconds with large clusters (due to the exhaustive search). On GPT-Neo, DTFM fails due to OOMs, while \sailor finds valid plans with a throughput of 0.07–0.21 iters/sec across cluster sizes. 

Finally, comparing \sailor's throughput for the OPT-350M model in the heterogeneous setups with more V100 (Figure \ref{fig:planner_het_OPT_25}) and the geo-distributed A100-only setup (Figure \ref{fig:planner_geo}), shows that the geo-distributed setup achieves up to 2$\times$ higher throughput, and also lower cost.

\textbf{\textit{Key takeaway 2:} Efficiently using the same GPU type across zones can lead to higher throughput and lower cost than mixing GPU types within a single zone, despite data transfer costs in geo-distributed setups. }

\subsubsection{Optimization with constraints}\label{sec:eval_constraints}

We now change the optimization objective to minimizing monetary cost and add constraints. Since baselines do not support cost-aware optimization or constraints, we modify them to rank solutions by iteration cost and only return plans within the constraints. We consider 2 cloud zones in the same region, each with 128 A100 and 128 V100. \sailor takes the full search space and outputs the resource allocation and parallelization plans. The baselines (that require a fixed topology) assume 4-GPU VMs: Varuna, Aceso, Galvatron  consider only the A100 machines (since they are more high-end than V100). AMP, FlashFlex, and Metis consider both A100 and V100 in a single zone, while DTFM considers only A100 in two zones. 

\begin{figure}[t]
    \centering
    \includegraphics[width=\linewidth]{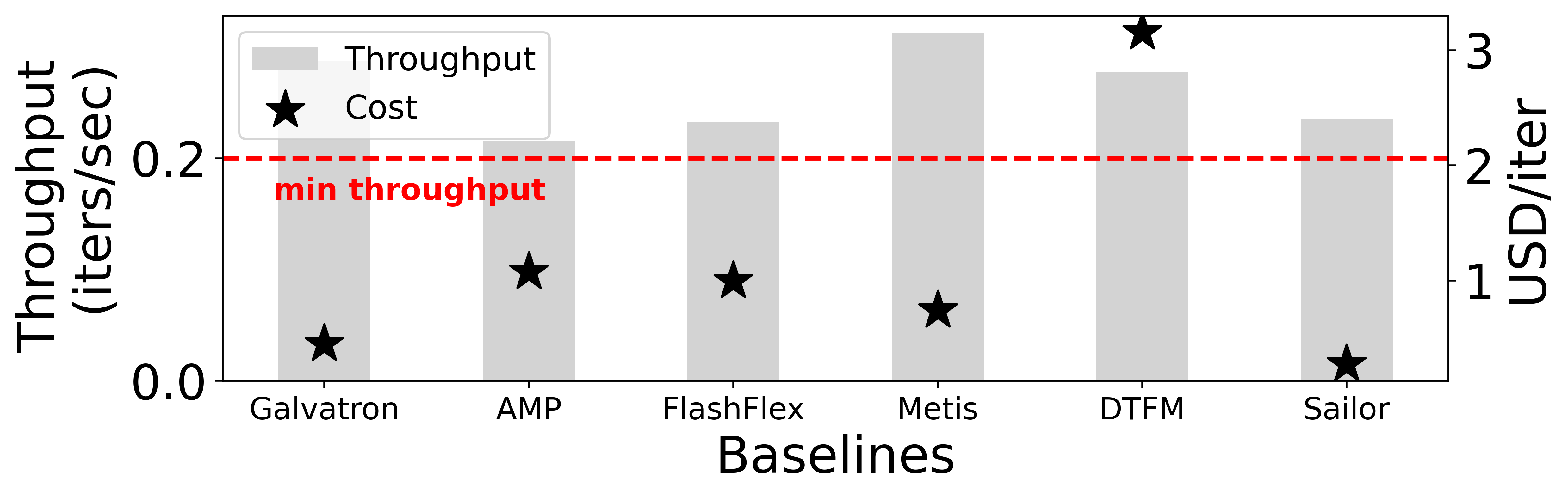}
    \caption{Minimizing cost with a throughput constraint.}
    \label{fig:plan_constraints_cost}
\end{figure}

\textbf{Scenario 1: Minimizing cost with throughput constraint of 0.2 iterations/sec:} Figure \ref{fig:plan_constraints_cost} shows the throughput (bars) and iteration cost (asterisks) achieved by the different planners. \sailor outputs a solution within the constraint, while achieving the minimum cost compared to baselines: 40\% lower cost compared to the second-best performing baseline (Galvatron). The found solution consisted of 64 A100 GPUs in a single zone, as they were enough to meet the throughput target.

\begin{figure}[t!]
    \centering
    \includegraphics[width=\linewidth]{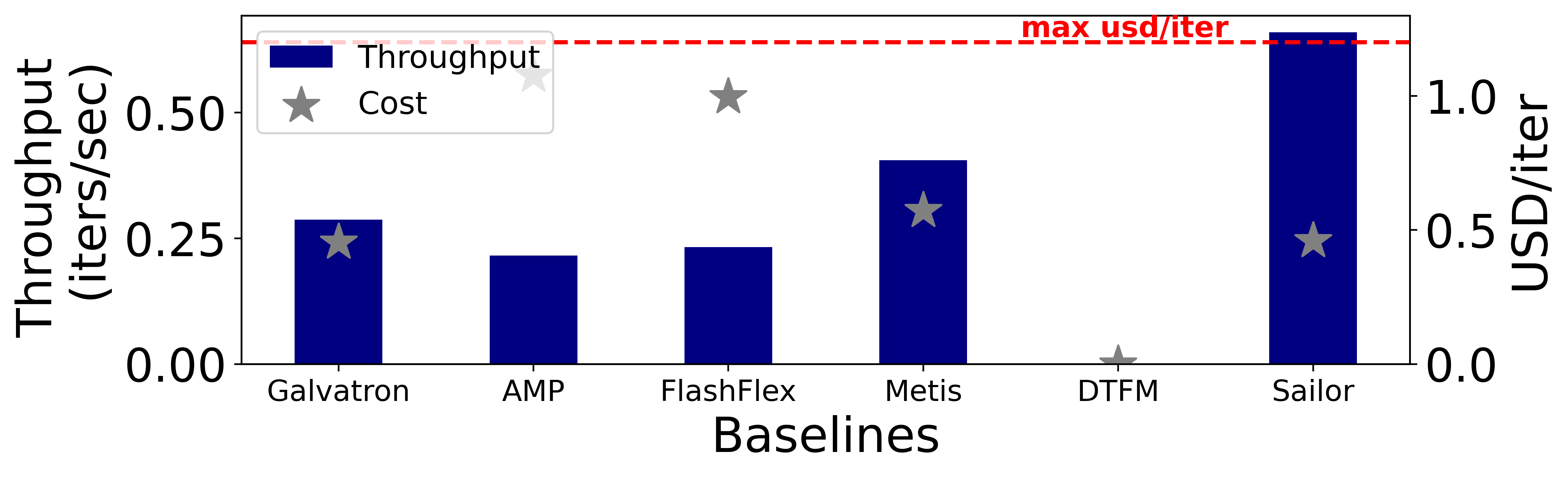}
    \caption{Maximizing throughput with budget limit.} 
    \label{fig:plan_constraints_thr}
\end{figure}

\textbf{Scenario 2: Maximizing throughput with cost constraint of 1.2 USD/iteration:}  Figure \ref{fig:plan_constraints_thr} shows the throughput (bars) and iteration cost (asterisks). Most baselines will use all available resources (e.g. all 128 A100 and 128 V100), even if they do not benefit throughput. DTFM does not find a solution as it outputs plans with low throughput and high costs. \sailor outperforms all baselines leading to 1.65-3$\times$ higher throughput while remaining within the cost constraint. \sailor's plan includes 256 A100 GPUs in two zones, with tensor parallelism of 4, and data parallelism of 64.

\subsection{\sailor scalability study}

We evaluated \sailor's search time varying the number of GPUs and zones or regions with a homogeneous GPU type. The search time remains below 1.5 sec even with 5 GCP zones and 256 A100/zone for the GPT-Neo-2.7B model. In contrast, adding more GPU types has a much higher impact in \sailor's search time: considering 256 GPUs/type in a single zone, \sailor's search time is 0.3, 6.2, 4900 sec for 1, 2, 3 GPU types, respectively. Nevertheless, \sailor's search process is much more efficient than the rest of the heterogeneous baselines:  Metis~\cite{Um24Metis} needs hours to complete its search even with 2 GPU types, while FlashFlex~\cite{yan2024flashflex} cannot find a valid configuration for any of these setups.

\subsection{\sailor Planner optimization breakdown} 

Table \ref{table:search_times_breakdown} shows \sailor's planner search time breakdown when optimizing for throughput, and budget constraint overhead, considering 128 A100 and 128 V100 GPUs for the GPT-Neo-2.7b model. The dynamic-programming-only approach needs hours to complete the search, even in the single-GPU-type case. Introducing the heuristics \textit{H1-H3} (\ref{planner_heuristics}), which apply in this scenario, dramatically decreases the search time to a couple of seconds.  The additional cost constraint increases the search time (4$\times$ increase in the 2-GPU-type scenario) due to the extra iterations caused by the straggler approximations, as described in \ref{planner_dp_cost}.

\begin{table}[tp!]
\centering
\begin{tabular}{|c|c|c|c|}
\hline
\textbf{GPU types} & \multicolumn{3}{c|}{\textbf{Search Time}} \\
\hline
& \textbf{Dyn Prog} & \textbf{+ Heuristics} & \textbf{+ cost limit} \\
\hline
\textbf{1} & hours & 0.25 sec & 0.4 sec \\
\hline
\textbf{2} & hours & 5 sec & 20 sec \\
\hline
\end{tabular}
\caption{Breakdown of search time, in dynamic programming and heuristics, and additional search time overhead due to budget constraints for the GPT-Neo-2.7 model. We use A100 and V100 in one zone, with 128 GPUs per type. The budget constraint is 1.5 USD/iteration.}
\vspace{-0.5cm}
\label{table:search_times_breakdown}
\end{table}

\subsection{Reconfiguration overheads}

We measure \sailor's reconfiguration time on a cluster of 16 V100 GPUs for the OPT-350M model. When 4 more GPUs are added, the controller re-invokes the planner, instructs workers to clean up (e.g., destroy NCCL groups), and broadcasts the new plan and topology. Planning takes 0.1 seconds, process cleanup takes 3 seconds, and topology broadcast (using grpc) takes 1.25 seconds. After the workers have received the new plan and topology, they reinitialize NCCL communication groups (for data, pipeline, tensor parallelism) in 4.5 seconds, redefine the model and optimizer in 2 seconds, redefine the dataloaders in 0.5 seconds, and resume training. 
While moderate at this scale, NCCL initialization can take minutes on thousands of GPUs~\cite{jiang2024megascale}. Existing methods to reduce this overhead~\cite{jiang2024megascale} can be integrated into \sailor.
\section{Discussion}

\textbf{Planner and simulator limitations: } Our planner and simulator currently support only the 1F1B pipeline parallel schedule, and do not incorporate optimizations such as activation offloading~\cite{Ren21zerooffload} or rematerialization~\cite{Yuan25remat}. Adding support for these optimizations is left for future work.

\textbf{Additional challenges with heterogeneous hardware:} Training over heterogeneous and geo-distributed datacenters can introduce additional challenges. Heterogeneity in accelerator vendors (e.g., NVIDIA vs. AMD) and network links (e.g., Infiniband vs. Ethernet) may prevent the use of high-performance collective communication libraries (e.g., NCCL), which often assume uniform network protocols. 
To maximize performance, collectives must be adapted to heterogeneous links. Furthermore, geo-distributed networks are prone to unreliability, including unpredictable jitter and packet loss. Training frameworks and algorithms should detect and adapt to such issues. Finally, even though \sailor optimizes parallelization strategies, it is based on strategies that were first introduced for homogeneous settings (e.g. 1F1B pipeline schedule). Achieving high performance in such contexts may require developing new schedules that more effectively overlap computation and communication, thereby reducing bubble times~\cite{chen2025crosspipe}.

\section{Other Related Work}

\textbf{Asynchronous geo-distributed training:} Several systems propose training over geo-distributed, preemptible, and heterogeneous resources by introducing asynchrony and reducing communication via quantization and sparsification. DiLoCo~\cite{douillard2024diloco} uses federated averaging to reduce communication, performing more local computation before synchronization. SWARM~\cite{ryabinin2023swarm}  proposes a decentralized, model-parallel approach to deal with poorly connected and unreliable devices. CocktailSGD~\cite{Wang23cocktailsgd} uses gradient compression to improve communication over low-bandwidth networks. These approaches influence training dynamics and are orthogonal to our work. \sailor employs synchronous training, which is preferred in large-scale training~\cite{wagenlander24tenplex}.

\textbf{Automatic VM selection}: CherryPick~\cite{Alipourfard17Cherrypick}, RAMBO~\cite{Li21Rambo}, and PARIS~\cite{Yadwadkar17Paris} apply Bayesian Optimization or performance modeling to recommend optimal VM types. SkyPilot~\cite{Zongheng23Skypilot} picks cost-efficient VMs across providers based on workload and user constraints. These methods treat workloads as black boxes and are not tailored for ML training. Srifty ~\cite{Luo22Srifty}, Cynthia~\cite{Zheng19Cynthia}, SpotDNN~\cite{Shang2023spotDNN}, DeepSpot~\cite{Lee17DeepSpotCloud} find optimal configurations of on-demand and spot instances for ML jobs. However, they target small data-parallel jobs, and are inadequate for large-scale training that involves various types of communication, which is our focus.

\textbf{Heterogeneous and Geo-distributed ML Inference:} Recent works have proposed systems for ML inference on heterogeneous and geo-distributed resources~\cite{Mei25Helix, Jian24Hexgen, Mao25Skyserve}. Although training and inference workloads differ significantly (e.g. in runtime and memory footprint estimation), synchronization and  communication patterns as well as planning decisions such as partitioning models across resources are relevant to both. Helix~\cite{Mei25Helix} and HexGen~\cite{Jian24Hexgen} consider both heterogeneous and geo-distributed resources for LLM inference. Helix uses a time-consuming MILP algorithm for model placement, which is not suitable for environments with high resource availability. HexGen uses a  more lightweight dynamic programming approach for splitting tensor and pipeline parallelism across resources, making it more appropriate for dynamic settings.
However, both of these works optimize only for performance, ignoring the monetary communication costs that arise in geo-distributed settings. SkyServe~\cite{Mao25Skyserve} places replicas of models across zones and regions, but restricts each replica to a single zone and homogeneous resources. As replicas do not communicate in inference, inter-zone communication is not considered. In contrast, \sailor targets geo-distributed training, which requires frequent communication among data-parallel replicas, substantially increasing scheduling complexity.

\section{Conclusion}

We propose \sailor, a system for efficient
large-scale training over heterogeneous resources with dynamic availability. \sailor co-optimizes the resource allocation and parallelization plan for a training job to optimize a user-defined objective, under constraints. By combining accurate iteration time and memory estimation, dynamic-programming based search, and domain-specific heuristics, \sailor efficiently navigates the large search space of possible job configurations. \sailor's distributed training framework supports heterogeneous setups and provides seamless elasticity. \sailor achieves 1.1-5.9$\times$ higher throughput than baselines across homogeneous, heterogeneous, and geo-distributed settings.
\section{Acknowledgements}

We thank the SOSP'25 anonymous reviewers and our shepherd, Ionel Gog, for their insightful feedback. We also thank Christina Giannoula for her feedback on the paper, and Michal Friedman for the helpful discussions.
This work was supported under project ID infra02 as part of the Swiss AI Initiative, through a grant from the ETH Domain and computational resources provided by the Swiss National Supercomputing Centre (CSCS) under the Alps infrastructure.
We thank the CSCS team for their technical support. Foteini Strati is supported by the Swiss National Science Foundation (project number 200021\_204620). George Manos is an Onassis Foundation scholar.

\newpage
\bibliographystyle{ACM-Reference-Format}
\balance
\bibliography{bibliography}

\end{document}